\documentclass[reprint,amsmath,amssymb,aps,superscriptaddress,nofootinbib,floatfix,12pt]{revtex4}
\usepackage{graphicx}
\usepackage{dcolumn}
\usepackage{bm}
\usepackage[utf8]{inputenc}
\usepackage{amsmath, amsthm, amsfonts, amssymb}
\usepackage[svgnames]{xcolor}
\colorlet{color1}{NavyBlue}
\usepackage[colorlinks=true,allcolors=color1]{hyperref}
\usepackage{physics}
\usepackage{dsfont}
\usepackage{cleveref} 
\usepackage{xfrac}
\usepackage{mathrsfs}
\usepackage[normalem]{ulem}
\usepackage{mathtools}

\begin{document}

\author{Julio Arrechea}
\affiliation{Instituto de Astrof\'isica de Andaluc\'ia (CSIC), Glorieta de la Astronom\'ia, 18008 Granada, Spain}

\author{Carlos Barcel\'o}
\affiliation{Instituto de Astrof\'isica de Andaluc\'ia (CSIC), Glorieta de la Astronom\'ia, 18008 Granada, Spain}

\author{Valentin Boyanov}
\affiliation{Departamento de F\'{\i}sica Te\'orica and IPARCOS, Universidad Complutense 
de Madrid, 28040 Madrid, Spain}
\affiliation{CENTRA, Departamento de F\'{\i}sica, Instituto Superior T\'ecnico, Universidade de Lisboa, Avenida Rovisco Pais 1, 1049 Lisboa, Portugal}


\title{After collapse: \\ On how a physical vacuum can change the black hole paradigm}

\begin{abstract}
Standard General Relativity assumes that, in the absence of classical matter sources, spacetime is empty.
This chapter considers and analyses the new behaviours of the gravitational field that appear when one substitutes this emptiness by a reactive vacuum, stemming in particular from the idea of vacuum provided by quantum field theory. 
We restrict our study to spherically symmetric configurations, and take a simple free quantum scalar field as a proxy to more complicated formulations.
Our analysis is split into a study of static and of dynamical configurations. 
Under the assumption of staticity, we find and describe the different asymptotically flat self-consistent solutions that appear when using a vacuum Renormalised Stress-Energy Tensor (RSET) as an additional source in the Einstein equations. 
Of particular interest is the discovery that, as opposed to standard general relativity, the new theory naturally contains static ultracompact stellar configurations which could observationally be mistaken for black holes (BHs).
Then, in our study of dynamical configurations, we investigate the possibility of these same vacuum effects changing the internal gravitational processes after an initial gravitational collapse in a way which shows a path towards forming the aforementioned ultracompact configurations. 
This has lead us to analyse several dynamical situations seldom contemplated in the literature. Of special relevance, we find that the inner horizon that all realistic BHs should contain could inflate outwards quickly enough to meet the outer one before any appreciable Hawking evaporation has taken place.
\end{abstract}
\maketitle

\section{Introduction}
\label{Sec:Introduction}

General Relativity (GR) is the best theory we currently have for describing gravitational phenomena. It is successful not only in terms of fitting observational data extremely well~\cite{Will2014}, but also due to the elegance of its formulation and the depth of its implications, which have given rise to mathematical and philosophical concepts that are still being developed to this day (see e.g.~\cite{Eisenstaedt2006,Brown2005}). However, it is generally accepted that GR is not the final theory of gravity. Much like Newtonian gravity before it, GR has limits in its range of applicability. This becomes apparent in problems which involve either strong gravitational fields on small scales (i.e. large densities and curvatures, or configurations involving black-hole horizons), or large-scale astrophysical and cosmological structures (i.e. the dark energy and, potentially, dark matter puzzles).

For strong fields in compact regions---this being the topic of the present chapter---the problem comes from the seemingly inevitable appearance of singularities when matter is compressed beyond certain limits~\cite{Penrose1965,SenovillaGarfinkle2015}, at least when this matter satisfies certain classical energy-positivity conditions~\cite{HawkingEllis1973}.
However, neither singularities themselves, nor a purely classical description of matter at such densities, can be considered physically reasonable. Indeed, the quantum nature of matter is generally thought to be directly related to the potential resolution of the singularity issue, along with a possible quantum behaviour of spacetime itself.

With these general considerations in mind, it is tempting to search for a new theory of gravity which is valid in these strong-field regimes. One possibility is to attempt to directly work out a quantum gravitational theory from first principles; examples of this are the Loop Quantum Gravity and String Theory programs. However, thus far these approaches have had major difficulties in making a clear connection between theory and phenomenology.  
A more conservative approach involves the search for improved theories of gravity that retain the idea of a classical spacetime geometry as an appropriate effective notion. In other words, maintaining an effective metric as one of the dynamical variables to be solved for, and only changing the dynamical equations it satisfies.

Taking the latter approach, however, suggests keeping in mind its potential limitations, given that a field theory with no pathologies whatsoever is an unlikely outcome. For instance, the Einstein field equations are rather unique in that they do not by themselves generate shockwaves or, in more generic terms, weak solutions (these only appear when the source fields themselves produce them~\cite{Reall2015}, or when a full-blown curvature singularity is approached~\cite{Marolf2012}). Modifying the Einstein field equations might achieve improved behaviours in certain situations, such as a removal of the formation of singularities in simple models of gravitational collapse, but this can potentially come at the cost of pathological behaviours in other situations. Within this effective spacetime perspective, the philosophy is therefore that of finding progressive improvements which are also compatible with previous successful theories, but not that of finding the \emph{final} dynamical theory for spacetime. Ultimately, the success of these improvements is then to be judged by how they accommodate the observed phenomenology.

The standard approach in modified gravity searches is to analyse sets of possibilities in theory space and classify them based on the phenomenology they present. The large variety of observed and measured gravitational phenomena then allows for the direct elimination of a substantial amount of these theories, while others have their free parameters constrained. Given the vastness of theory space, to even begin such an analysis calls for some formal or physical arguments which allow for a selection of a specific theory or set of theories. Different researchers have different tastes and criteria for such a selection, resulting in the dendritic exploration of theory space currently being carried out~\cite{Olmo2011,Petrov2020,BohmerJensko2021}.

New theories can be broadly categorised according to two non-mutually-exclusive structural features: a) those which modify how matter behaves in strong gravity situations while retaining the form of the Einstein equations, or b) those which modify directly the gravitational equations, without necessarily modifying the matter sources. Within the latter set, proposals can be found that change the geometry significantly only in the surroundings of would-be classical singular regions, and others that can lead to alterations of the geometry well beyond such regimes. There are numerous interesting proposals out there, some of them discussed in the chapters of this book.

This chapter in particular is centred around one modification of GR which can be considered as part of category b). The idea is the following: we assume that the zero-point fluctuations of the quantum fields permeating spacetime gravitate by some amount determined by the very deviation from flatness of the spacetime, generally in a non-local manner (we note that at this stage we will neglect the possible presence of a global contribution in the form of a cosmological constant); then, even in regions where no classical matter exists, there would be an average vacuum energy (as well as pressure, fluxes, etc.) which would be a source of gravity. As this new source ultimately depends on the geometry, we can interpret it as part of the geometric side of Einstein equations, maintaining that the only real source on the right-hand side is the classical stress-energy tensor. An alternative way of describing this approach is by saying that the standard vacuum used in GR is, in a sense, ``too empty"; this approach instead aims to characterise a more physical vacuum as the stage on which all gravitational phenomena take place.
Before we describe this theory in more detail, let us first make a brief digression and discuss the notion of a theory of quantum gravity.

It is clear that Nature has a way of melding together both the quantum behaviour that we observe for matter in the atomic and subatomic regimes and the gravitational behaviour that we observe in the macrocosm. A theory of quantum gravity would be a model that consistently incorporates both of these behaviours, furthering our understanding of each. 
Thus far, we have had no clear indication of how these two regimes come together; however, theoretical considerations do provide some clues. For instance, in the direction of {\it how gravity affects matter} there are strong indications that the causality provided by a geometric gravitational description affects the behaviour of quantum matter (see for example~\cite{Mulleretal2010}). Under this hypothesis, analyses of quantum field theory over fixed curved background spacetimes have lead to some of the most important results in modern theoretical physics. On the one hand, the idea that a quantum particle is an observer-dependent notion leads to analysis of particle production in cosmology~\cite{Parker1968,Fulling1973}, and subsequently to the idea of how an inflationary regime could lead to a primordial spectrum of fluctuations in the early universe~\cite{Starobinsky1979,MukhanovChibisov1981}. On the other hand, a calculation of particle production associated to the formation of a black hole (BH) lead Hawking to his famous result that BHs should evaporate~\cite{Hawking1974,Hawking1975}.

In the opposite direction of {\it how matter affects gravity}, GR tells us that the average effect of the stress-energy contained in a macroscopic lump of matter is to bend causality in particular, well-defined ways. However, we know nothing of how a single quantum lump of matter affects causality, or even whether a classical notion of causality would be valid at these scales. In fact, it would not be surprising if it turned out that gravity as we know it is not a relevant notion until matter starts behaving classically due to its aggregation~\cite{Penrose1996,Penrose2014}. Indeed, such regimes are where GR is less known and less clearly tested, to say nothing of when matter enters into full Planckian-density regimes. Putting aside the precise resolution of these issues, it is at least reasonable to believe that there is a regime in which an effective classical spacetime makes sense, with curvature being sourced by the average energy contained in quantum states. Then, aside from the intrinsic excitations of quantum fields, which we usually define as matter, one must also take into account the energy which can be generated by the very presence of curvature in the spacetime. Since on large scales matter can, for the most part, be described classically, in this regime the total stress-energy tensor in the Einstein equations should contain a classical contribution, where actual matter is localised, plus another source term taking into account the average energy of the vacuum fluctuations (as mentioned before, this last term can also be moved to the geometric side of the equations).
This description would be self-consistent whenever the fluctuations of the energy around the average (or expectation value) are small~\cite{HuVerdaguer2020}, as one might expect to be the case for a vacuum state in standard situations, or in quasi-classical states.

When performing calculations within standard relativistic quantum field theory in Minkowski spacetime, there are no observables that couple to an absolute notion of energy---they only couple to differences of energy between different states. Thus, one can always subtract from all energy measurements an arbitrary reference value. In fact, a first naive calculation indicates that any quantum state has an infinite energy, as is shown in any introductory text on the subject~\cite{PeskinSchroeder1995}. However, there is the freedom to renormalise the energy of all states in such a way that the lowest energy state, i.e. the vacuum state, has zero energy. By subtracting the ``same" infinite value from the energy of all states, we obtain finite definite values for finite-particle states (with respect to the chosen zero point). This can be done easily in flat spacetime because there is a unique natural notion of a vacuum state (the so-called Minkowski vacuum), and subtracting its expectation values from other states is a procedure which preserves the symmetry of the background.

However, the theory changes in an essential way in the presence of spacetime curvature, as there is no longer a natural notion of vacuum state. Indeed, extensions of the flat spacetime theory only lead to conclude that different observers can have a different perception of what the true vacuum state is. There is generally no longer a ``vacuum" state which can be used for renormalisation by subtraction, as such a procedure would break the symmetries of GR. However, given that gravity is sensitive to the total energy of a state, rather than just energy differences, renormalisation is of particular importance. Extreme care is required in selecting how and what to subtract in search of sensible results.

With these observations in mind, the requirements for renormalisation indeed change quite a bit. Particularly, one preferably needs a local subtraction prescription which does not make use of knowledge of either the global structure of spacetime, or of the particular vacuum and particle states chosen for quantisation. Prescriptions of this sort do exist, but they generally leave a non-homogeneous, finite vacuum energy residue, which depends on the characteristics of the spacetime and of the chosen vacuum state~\cite{BirrellDavies1982,Wald1994}.
The subtraction can be made consistent with recovering asymptotic Minkowski spacetime when going arbitrarily far from the classical sources of curvature, or it can be made to leave an offset in the form of a cosmological constant.
As in the present chapter we are not considering cosmological scenarios, we will neglect such terms, and set the cosmological constant to zero.

The central hypothesis of this chapter is that the theory of classical matter plus vacuum fluctuations sourcing gravity is applicable in the astrophysical scenarios we analyse below. Formally, this theory is constructed with the effectively classical stress-energy tensor (SET), and the residual vacuum energy, or zero-point fluctuations, of the quantum fields as sources in the Einstein field equations,
\begin{eqnarray}\label{Eq:SemiEqs}
G_{\mu\nu}=8\pi G( T^{\rm C}_{\mu\nu}+ T^{\rm ZP}_{\mu\nu} ),
\end{eqnarray}
were $T^{\rm C}_{\mu\nu}$ is the classical and $T^{\rm ZP}_{\mu\nu}$ the zero-point stress-energy source term.
Equivalently, from a modified gravity perspective the equations can formulated as
\begin{eqnarray}
G_{\mu\nu}- 8\pi G\,T^{\rm ZP}_{\mu\nu} = 8\pi G\,T^{\rm C}_{\mu\nu}.
\end{eqnarray}
These formal equations of motion constitute what is commonly referred to as \textit{semiclassical gravity}. Indeed, the presence of a non-trivial zero-point SET is one of the most soundly motivated modifications of gravity we currently have. Though we have no direct observational evidence of this gravitational vacuum polarisation, we can straightforwardly make an analogy with the case of electromagnetism, where we know that a similar notion of charge polarisation does exist~\cite{LambRetherford1947}.

Up to this point we have talked about vacuum energy in a deliberately vague manner.
If we knew the exact expression of $T^{\rm ZP}_{\mu\nu}$ in terms of simple analytic functions of the metric, we would just have to solve the new system of gravitational equations and analyse its phenomenology, testing against observations. 
However, things are not so simple, as we do not possess an indisputable prescription for calculating $T^{\rm ZP}_{\mu\nu}$ in generic scenarios. The best reasoned method we have for obtaining an appropriate $T^{\rm ZP}_{\mu\nu}$ is through the expectation value of a SET operator for different fields in a given vacuum state,
\begin{eqnarray}
T^{\rm ZP}_{\mu\nu} = \langle \Psi_0 | \hat{T}^{\rm QF}_{\mu\nu} |\Psi_0 \rangle.
\end{eqnarray}

Ideally, $T^{\rm ZP}_{\mu\nu}$ would describe the SET operator associated with the complete standard model of particle physics, with all its interacting fields, together with any yet undiscovered fields, such as the possible constituents of dark matter. Additionally, $T^{\rm ZP}_{\mu\nu}$ would incorporate any fluctuating energy offset that might be contained within the gravitational field itself. Neither of these idealised requirements is realistically feasible as of yet. On the one hand, given the absence of a theory of quantum gravity, we are bound to hope that the gravitational contribution to these fluctuations is small enough to be negligible with respect to the ones associated to the standard model fields\footnote{It has been suggested that this assumption becomes more accurate when the number of quantum fields $N$ is sufficiently large~\cite{Anderson1983,BirrellDavies1982}}.
On the other hand, even the standard model zero-point SET is nearly impossible to calculate. Such difficulty arises from two fundamental aspects:
a) it is unknown how to treat interacting field theories beyond the $S$-matrix perturbative approach; b) as the SET operator contains products of field operators at the same point, it is not a well defined operator in the quantum theory; therefore, to make sense of this object one has to resort to regularising and renormalising this expectation value, which poses its own difficulties in curved spacetimes. Faced with these problems, as a proxy to the qualitative form that $T^{\rm ZP}_{\mu\nu}$ might have, researchers have opted for calculating the renormalised SET (RSET) for free field theories (often as a one-loop approximation to interacting theories) in simple backgrounds. Among the test fields useful to understand semiclassical gravity, the free scalar field is the simplest, and indeed the most used in past and present literature. The term \textit{semiclassical gravity} is typically used to refer to any theory that incorporates the effect of vacuum fluctuations, even in these simplified test field scenarios.

Even calculating the RSET of a free scalar field in simple, highly-symmetric geometries is not a trivial task; in fact, it can typically only be done numerically and with great difficulties. The complete problem of solving the semiclassical Einstein equations self-consistently and exactly is therefore not feasible at present.
This problem has a large body of work addressing it, and we will briefly and non-exhaustively review it in the next section.
For now let us just say that given the difficulties mentioned, there are essentially two strategies that can be adopted: 1) to develop incremental improvements in the method of calculating the RSET, or very close approximations thereof, and study its effects in physically relevant situations in an approximate perturbative manner; 2) to prescribe less precise but analytically simpler approximations to the RSET, such that one can readily investigate more complicated and realistic geometric situations, understanding meanwhile that the information extracted only provides a qualitative preview of what a full exact solution may look like. In the following sections of this chapter we present an application of the second strategy to gravitational collapse and compact object geometries.

Particularly, we will present work on two closely related physical problems. Firstly, we will address the question of whether semiclassical gravity allows for qualitatively different configurations of stellar equilibrium as compared to those present in GR. Then, we will look into the problem of gravitational collapse with a bit more scrutiny under a semiclassical lens. In other words, we will present results for both (meta-)stable configurations of semiclassical gravity which allow the existence of stellar objects of higher compactness than their classical counterparts~\cite{Arrecheaetal2021c,Arrecheaetal2022}, as well as for a revised dynamical process of gravitational collapse which may lead to their formation~\cite{Barceloetal2019,Barceloetal2020,Barceloetal2022}.  

Even considering the approximations and hypotheses involved, our philosophy with these analyses is to progressively build up and develop a modified gravitational model which is consistently treatable and comparable with GR on an equal footing. In this endeavour, both of the above-mentioned strategies can greatly contribute. While such a theory would by no means be a complete and exact treatment of matter at high densities (due to the approximations involved, as well as due to neglecting the fact that even classical matter at such densities would likely behave differently from what is known), it is our best attempt to push gravity onto the next stage of development.

The outline of this contribution is the following. In the next section we briefly review the status of research into calculating the RSET. Then, section~\ref{Sec:Approximations} will describe two qualitative approximation to the RSET which are analytically manageable: the Regularised Polyakov approximation and the Order-Reduced Anderson-Hiscock-Samuel tensor. Armed with these two RSETs, in section~\ref{Sec:Static} we will present an analysis of how the vacuum-induced changes in the equations of hydrostatic equilibrium lead to new families of stellar solutions absent in GR. Subsequently, in section~\ref{Sec:Dynamical} we will analyse the dynamics of gravitational collapse, paying close attention to possible modifications to the standard picture due to horizon-related effects. Finally, we will summarise our findings and conclude with some final remarks.

\section{Semiclassical gravity: in search of an appropriate RSET}
\label{Sec:RSET}

Within the semiclassical approach, the central and most important problem is the search for non-ambiguous and feasibly calculable RSETs. At the present stage, calculations are usually performed for simple test fields, as they provide an invaluable glimpse into the potential behaviours hidden within the semiclassical theory. Let us consider in particular the free scalar field. This is typically used in the literature~\cite{HuVerdaguer2020} as the starting point for considering more complicated fields and interactions, which, while bringing about additional contributions, are not expected to lead to fundamental changes in the vacuum dynamics of semiclassical systems. For instance, particle creation processes in cosmology and gravitational collapse occur in a qualitatively robust way for a variety of different fields~\cite{Ford2021,BirrellDavies1982,Page2004}, showing that test-field semiclassical analyses suffice for a qualitative analysis, and even for some quantitative estimates~\cite{Hawking1974}. Hereafter, the tool used for the entirety of the discussion will be the scalar field.

In general terms, we can define the RSET of a scalar field in some vacuum state as the result of applying a regularisation and renormalisation procedure ${\cal P}$ to the ill-defined object $\langle 0 | \hat{T}_{\mu\nu} | 0 \rangle$. Symbolically we can write
\begin{eqnarray}
\langle\hat{T}_{\mu\nu}\rangle_{\text{ren}} := {\cal P} \left[ \langle 0 | \hat{T}_{\mu\nu} | 0 \rangle \right]
\end{eqnarray}
Several regularisation techniques have been developed in the literature to find expressions for $\langle\hat{T}_{\mu\nu}\rangle_{\text{ren}}$. Examples of that are: covariant point separation (or point-splitting)~\cite{Christensen1976,Christensen1978}; Hadamard regularisation~\cite{BrownOttewill1985}; dimensional regularisation~\cite{BolliniGiambiagi1972}; Riemann $\xi$-function regularisation~\cite{Hawking1976}; Pauli-Villars regularisation~\cite{PauliVillars1949};  proper time~\cite{Schwinger1951} and adiabatic regularisation~\cite{Parker2009}. The choice of an appropriate method depends on the particularities of the system under consideration, but the underlying logic is common in all of them: the subtraction of local divergences is carried out in a way which preserves general covariance.
Under some reasonable assumptions, it turns out that different regularisation procedures lead to RSETs that differ at most in locally constructed conserved quantities, as proven by Wald~\cite{Wald1977,Wald1978}. After the divergent terms have been subtracted, one can find a well-defined RSET with the appropriate physical characteristics expected from a source term in the Einstein equations~\cite{Wald1994}.

The task of identifying divergent terms in generic spacetimes so that they can be subtracted was completed by the end of the seventies~\cite{Christensen1978}.  
Nonetheless, the problem of how to calculate the remaining finite terms in the most efficient and accurate manner has remained. The expectation values involved are constructed through a spectral decomposition of the field, as quantisation and the definition of a vacuum state themselves rely on choosing and obtaining a specific basis of modes which satisfy the field equation of motion. It is at this stage where we encounter the main difficulty of the problem: field modes cannot be calculated in closed form for most of the spacetimes of interest, let alone for generic spacetimes.

Given this situation, the approaches used for calculating the RSET are split into two categories: on the one hand, there are those which look for appropriate approximation schemes that are as accurate as possible, and on the other, those which implement progressively more efficient numerical schemes. For instance, the first methods used to calculate the RSET in Schwarzschild spacetime made use of Wentzel–Kramers–Brillouin (WKB) approximations to the modes. The WKB based method was originally devised in~\cite{Candelas1980,Howard1984,HowardCandelas1984}, and has been improved over the years~\cite{Anderson1989}. The epitome of applying this method can perhaps be found in the work by Anderson-Hiscock-Samuel~\cite{Andersonetal1995}. They were able to give expressions for the RSET in arbitrary static, spherically-symmetric spacetimes in the Hartle-Hawking vacuum state (there are results also for the Boulware vacuum state in~\cite{Levi2016}) for a scalar field with arbitrary mass and coupling to curvature. The RSET they found is split into two terms, each of which is conserved separately. The first has an analytic closed form, and the second is in general only obtainable numerically. Additionally, they showed that for massless fields the analytic part on its own constitutes a good approximation to the total RSET. In fact, for the conformally invariant field this approximate RSET was found before by Page~\cite{Page1982} and later on by Frolov and Zelnikov~\cite{FrolovZelnikov1987}.

However, this analytical approximation to the RSET, which we will refer to as the AHS-RSET, has a number of shortcomings for its implementation in the field equations of gravity. Chief among them is the fact that it depends on up to fourth order derivatives of the metric functions. Firstly, this makes the solutions of the semiclassical Einstein equations depend on too many boundary conditions, with no clear physical interpretation (such as initial momentum, reflective boundaries, etc.). Secondly, when analysing self-consistent solutions of the system of equations, spurious solutions connect to seemingly tame initial conditions, making it so that even Minkowski spacetime can be unstable under small perturbations, as was found even before this approximation to the RSET was obtained~\cite{HorowitzWald1980}. This is reminiscent to what occurs with the Lorentz-Abraham-Dirac self-force of classical electrodynamics, which contains self-accelerated solutions~\cite{Rohrlich1999,Rohrlich2000}. In that case we can track the origin of these solutions and improve the system of equations introducing some integro-differential operator~\cite{Rohrlich1999}, or applying an order-reduction procedure~\cite{Simon1990,FlanaganWald1996}. 

Another issue with the AHS-RSET is that the WKB expansion it relies on breaks down at the horizon of BH geometries. This results in the approximate RSET having logarithmic divergences at the horizon, present even in the supposedly regular Hartle-Hawking vacuum state. These divergences have been shown to disappear when one adequately adds the numerical part of the tensor, as outlined in~\cite{Andersonetal1995} (alternatively, one can also deal with them by treating the first modes of the expansion separately~\cite{Balbinotetal2000}). However, these issues make it so that using the AHS-RSET directly to find self-consistent solutions is rather complicated and the results are somewhat untrustworthy. 
Nonetheless, certain physical scenarios do allow for the approximation as such to be useful, such as for the wormhole solutions found in~\cite{Hochbergetal1997}.

In more recent times, two new methods for obtaining the RSET have been devised. One is the so-called {\em pragmatic mode-sum method} pioneer by Levi and Ori~\cite{LeviOri2015}. The other is the  {\em extended-coordinates method} proposed by Taylor and Breen~\cite{TaylorBreen2016}. The first is a completion and generalisation of a method developed by Candelas~\cite{Candelas1980}. It does not make use of WKB expansions in the corresponding Euclidean sector. In fact, it does not use WKB expansions at all, since in the Lorentzian sector high-order WKB approximations are very cumbersome to use, as they involve using asymptotic matching techniques to approximate the modes at turning radii at which the approximation breaks down. Instead, the idea is to construct generalised integrals in frequency which directly incorporate a subtraction of the divergences (based on high frequency information in Christensen counterterms~\cite{Christensen1978}), in such a way that the convergence of the integral is efficient. The advantage of this method is that it can be applied to dynamical situations (e.g. Hawking evaporation) and also to axisymmetric configurations (e.g. in~\cite{Levietal2016} it was used to calculate the RSET on a Kerr background).

The extended coordinate method~\cite{TaylorBreen2016,TaylorBreen2017} is performed in the Euclidean sector, and so it is not suitable for dynamical configurations. It introduces some new useful coordinates to decompose the Hadamard parametrix into multipoles and Fourier components. This allows for a mode by mode subtraction in a manner that results in a numerically very efficient algorithm. Summing up a few tens of modes gives quite accurate results and the method can be applied equally well to higher dimensional spacetimes~\cite{BreenOttewill2012,Morleyetal2021a,Morleyetal2021b,Tayloretal2022}.

These developments are part of an exciting progress trend and motivated on excellent grounds. Due to recent computational advances, it is expected that progress in the efficiency of calculating the RSET in scenarios of greater phenomenological interest will carry on. Nonetheless, it is difficult to foresee how these RSETs could be used to search for self-consistent solutions. The main obstruction here owes to the complexity of simultaneously finding the geometry of spacetime and the field modes propagating on (and being sources of) that very same spacetime\footnote{In this regard, we could aim for developing efficient grid-search algorithms that converge to a metric satisfying the semiclassical equations, but even this possibility seems to us computationally discouraging.}. For this reason, we believe it is worth considering approximations to the RSET which, despite being less accurate, prove better suited for finding self-consistent solutions in semiclassical gravity.

\section{Approximate RSETs}
\label{Sec:Approximations}

Analytic approximations to renormalised stress-energy tensors are all based on a similar rationale of finding trade-offs between the functional complexity of these RSETs and the accuracy of the physics they encode. As mentioned above, we will use a free scalar field throughout the remainder of the chapter. This field obeys the equation of motion
\begin{eqnarray}\label{Eq:WaveEquation}
    \Box\phi-\left(m^2+\xi R\right)\phi=0,
\end{eqnarray}
where $\Box$ is the d'Alembertian operator, $m$ is the field mass, and $\xi$ the coupling to the Ricci scalar $R$. A commonly used method for obtaining analytic, approximate expressions for the RSET is based on fixing the field parameters (the mass and curvature coupling) and restricting analyses to particularly simple spacetimes. For instance, for conformally invariant fields \mbox{$(m=0,~\xi=1/6)$} on conformally flat backgrounds, the RSET is determined by the local trace anomaly~\cite{Page1982, BrownOttewill1985}. This leads to the existence of explicit analytic expressions for the RSET in a variety of situations, such as Friedmann-Lemaître-Robertson-Walker cosmologies and stellar interiors of constant density~\cite{ParkerSimon1993}~(for the particular quantisations which are formulated in accordance with the conformal symmetry), and even for fields of higher spin~\cite{FrolovZelnikov1987}.

However, for the situations we will analyse below, such as the formation and subsequent evaporation trapped regions or the existence of ultracompact stellar configurations in equilibrium, spacetime is not conformally flat. In such cases the RSET includes contributions which are non-local in curvature, and that depend on the vacuum state under consideration.
The simplest RSET that captures these state-dependent effects in spherical symmetry is the Polyakov approximation, which incorporates the essential features of the propagation of a massless minimally coupled scalar $(m=0,~\xi=0)$ in four spacetime dimensions via two-dimensional model, described by the line element
\begin{eqnarray}\label{eq:2Dmetric}
ds^{2}_{\text{(2D)}}=-\mathcal{C}(u,v)dudv,
\end{eqnarray}
where $u=t-r^{*},~v=t+r^{*}$ are radial null coordinates, with $r^{*}$ the tortoise coordinate obtained by integrating $dr^*/dr=[h(r)/f(r)]^{1/2}$. The analogy between scalar field propagation in four and two spacetime dimensions becomes clear when one expands the field in spherical harmonics and restricts the analysis to the $l=0$ (or $s$-wave) mode, which typically dominates long-distance effects. By considering the propagation of the $s$-wave over BH spacetimes and taking the near-horizon limit in the wave equation~\eqref{Eq:WaveEquation}, the part of the equation that can be identified as a gravitational potential vanishes, and the $(t,r)$ sector reduces to the two-dimensional free wave equation
\begin{eqnarray}\label{Eq:WaveEq2D}
    \partial_{u}\partial_{v}\phi=0.
\end{eqnarray}
This equation is manifestly conformally invariant, and admits an analytic basis of solutions in the form of plane waves. In fact, there are infinitely many such bases, one for each pair of possible null coordinates, and each can be used for performing quantisation~\cite{Barceloetal2012}. Using any one of these quantisations, the two-dimensional RSET can be obtained in closed analytic form~\cite{DaviesFulling1977}, in particular,
\begin{align}\label{Eq:2DRSET}
    \langle\hat{T}_{uu}\rangle^{\text{(2D)}}=
    &
    \frac{1}{24\pi}\left(\frac{\mathcal{C}_{uu}}{\mathcal{C}}-\frac{3\mathcal{C}_{u}^{2}}{2\mathcal{C}^{2}}\right),\nonumber\\
    \langle\hat{T}_{vv}\rangle^{\text{(2D)}}=
    &
    \frac{1}{24\pi}\left(\frac{\mathcal{C}_{vv}}{\mathcal{C}}-\frac{3\mathcal{C}_{v}^{2}}{2\mathcal{C}^{2}}\right),\nonumber\\
    \langle\hat{T}_{uv}\rangle^{\text{(2D)}}=
    &
    \langle\hat{T}_{vu}\rangle^{\text{(2D)}}=-\frac{R^{\text{(2D)}}}{96\pi}\mathcal{C},
\end{align}
where $R^{\text{(2D)}}$ is the two-dimensional Ricci scalar. The difference between the results obtained for the modes corresponding to one pair of null coordinates or another, i.e. between different choices of vacuum state, comes in the form of $uu$ and $vv$ flux terms,
\begin{align}
\langle\hat{T}_{uu}\rangle^{\text{(2D)}}&\xrightarrow[\text{vac.~change}]{}\langle\hat{T}_{uu}\rangle^{\text{(2D)}}+\langle:\hat{T}_{uu}:\rangle,\nonumber\\
\langle\hat{T}_{vv}\rangle^{\text{(2D)}}&\xrightarrow[\text{vac.~change}]{}\langle\hat{T}_{vv}\rangle^{\text{(2D)}}+\langle:\hat{T}_{vv}:\rangle,\nonumber\\
\langle\hat{T}_{uv}\rangle^{\text{(2D)}}&\xrightarrow[\text{vac.~change}]{}\langle\hat{T}_{uv}\rangle^{\text{(2D)}}
\end{align}
The terms $\langle:\hat{T}_{\mu\nu}:\rangle$ incorporate all the dependence on the state in the RSET. For instance, in the case of BHs, when one switches between the Boulware and Unruh states, they contain the fluxes across horizons responsible for the phenomenon of Hawking evaporation ~\cite{FabbriNavarro-Salas2005}. They are obtained through the Schwarzian derivative between the null coordinates which encode the different quantisations~\cite{Fabbrietal2003,FabbriNavarro-Salas2005}.

Having chosen a particular quantisation and obtained the two-dimensional RSET, the next step in the Polyakov approximation consists in defining a four-dimensional RSET from the components~\eqref{Eq:2DRSET} through the relations
\begin{eqnarray}\label{Eq:2Dto4D}
\langle\hat{T}^{\mu}_{\nu}\rangle^{\rm P}=F(r)\delta^{\mu}_{a}\delta^{b}_{\nu}\langle\hat{T}^{a}_{b}\rangle^{\text{(2D)}}
+(T_{\rm AC})^{\mu}_{\nu},
\end{eqnarray}
where Greek and Latin indices take $4$ and $2$ values, respectively, $\rm P$ stands for Polyakov RSET, $F(r)$ is a radial function that up-scales the tensor to four dimensions, and $(T_{\rm AC})^{\mu}_{\nu}$ is a term which encodes angular pressures not contained in the two-dimensional theory. The standard Polyakov approximation is obtained with the choice $F(r)=1/4\pi r^{2}$, which mimics the four dimensional behaviour of spherical modes, and $(T_{\rm AC})^{\mu}_{\nu}=0$. This already suffices to reproduce the behaviour of the RSET at BH horizons~\cite{FabbriNavarro-Salas2005,DFU}, and is generally expected to work well far away from the origin.

The Polyakov RSET has been extensively used to study the semiclassical backreaction problem in a variety of situations, from dynamical BH formation and evaporation to static stars in equilibrium~\cite{Chakraborty2015,ParentaniPiran1994,Fabbrietal2006,Carballo-Rubio2018}. The fact that this RSET is analytic, simple and has only up to second derivatives of the metric functions allows one to pose an evolution problem that is not too different from that of general relativity. 

However, due to the divergence at the origin for this standard choice of $F(r)$, one can instead use a regularised version of it when dealing with systems which include $r=0$. On its own, this comes at the price of breaking conservation, but this can be compensated by the introduction of an appropriate $(T_{\rm AC})^{\mu}_{\nu}$ term. We will refer to this tensor as the \emph{regularised} Polyakov RSET (RP-RSET), which we will use for stellar configurations in the next section. The $F(r)$ function can be fixed freely in static configurations (though the $1/r^2$ form should be retained far away from the origin in order to recover the s-wave behaviour), as the conservation equations
\begin{align}\label{Eq:ConsRSET}
    \nabla_{\mu}\langle\hat{T}^{\mu}_{r}\rangle^{\rm P} =
    ~\partial_{r}\langle\hat{T}^{r}_{r}\rangle^{\rm P} +\frac{2}{r}\left(\langle\hat{T}^{r}_{r}\rangle^{\rm P} -\langle\hat{T}^{\theta}_{\theta}\rangle^{\rm P} \right)
    +\frac{f'}{2f}\left(\langle\hat{T}^{r}_{r}\rangle^{\rm P} -\langle\hat{T}^{t}_{t}\rangle^{\rm P} \right)=0,
\end{align}
can be satisfied with an appropriate choice of $(T_{\rm AC})^{\mu}_{\nu}$. In dynamical situations, however, such a regularisation is not as straightforward, and one needs a more thorough deformation of the RSET components near the origin. In the following we only use the RP-RSET in static scenarios, while in dynamical ones we simply steer clear of the origin for now, focusing instead on the vicinity of horizons.

For completeness let us also mention that it is possible to incorporate into the Polyakov RSET the backscattering effects of the gravitational potential by considering a two-dimensional scalar coupled to a dilaton field (we refer the reader to~\cite{Fabbrietal2003,FabbriNavarro-Salas2005} for details on this approach). This method has a similar issue to the RP-RSET, in the sense that the resulting two-dimensional RSET is not conserved; this is again compensated with angular components in the four-dimensional tensor. We avoid the use of this approach in the following analyses, since the RSET it gives exhibits similar problems as the standard Polyakov RSET at $r=0$, while at the same time containing terms with third order spatial derivatives of the metric functions.

The simplicity of the RP-RSET grants it a sort of malleability that makes it applicable to a variety of static scenarios. The static and spherically-symmetric geometries we will work with have a metric which can be written as
\begin{eqnarray}\label{Eq:4DMetric}
    ds^{2}=-f(r)dt^{2}+h(r)dr^{2}+r^{2}d\Omega^{2}.
\end{eqnarray}
Leaving $F$ unspecified and choosing the quantisation which respects staticity (corresponding to the Boulware vacuum), we arrive, through~\eqref{Eq:ConsRSET}, at the components of the RP-RSET,
\begin{align}\label{Eq:PolyakovRSET}
    \langle\hat{T}^{t}_{t}\rangle^{\rm P}=
    &
    \frac{F}{96\pi h}\left[\frac{2f'h'}{fh}+3\left(\frac{f'}{f}\right)^{2}-\frac{4f''}{f}\right],\nonumber\\
    \langle\hat{T}^{r}_{r}\rangle^{\rm P}=
    &
    -\frac{F}{96\pi h}\left(\frac{f'}{f}\right)^{2},\nonumber\\
    \langle\hat{T}^{\theta}_{\theta}\rangle^{\rm P}=
    &
    -\frac{\left(2F+rF'\right)}{192\pi h}\left(\frac{f'}{f}\right)^{2}.
\end{align}
This will be the RSET we will use for the majority of the next section, where we analyse self-consistent semiclassical solutions. Before this, let us make some final remarks on other existing RSET approximations in static spacetimes.

Approximate RSETs based on dimensional reduction have been extensively used despite their problematic behaviour at $r=0$. These problems are absent from RSET approximations that consider the four-dimensional field dynamics from the start, like the analytic approximation derived by Anderson, Hiscock and Samuel~\cite{Andersonetal1995}. Although this approximation gives a well-behaved RSET at $r=0$, the AHS-RSET has additional problems for massive fields. For example, in flat spacetime it contains non-zero contributions that cannot be renormalised away~\cite{Arrecheaetal2023} nor be identified with a cosmological constant~\cite{Martin2012}.

Some of the complications associated with the AHS-RSET can be circumvented by considering just the zero-mass case and applying to it a reduction-of-order procedure (see~\cite{ParkerSimon1993,Arrecheaetal2023}). After the reduction the resulting system of differential equations is again second order a thus adequate for backreaction analyses~\cite{Arrecheaetal2023,Arrechea2023}. In the next section we will use mostly the Regularised Polyakov approximation, but also this Order-Reduced AHS-RSET to find self-consistent solutions of the semiclassical equations both for classically empty spacetimes and for stars of constant density.

\section{Stellar equilibrium on a physical vacuum soil}
\label{Sec:Static}

Equipped with a suitable RSET satisfying the desired properties of analyticity, regularity, conservation and low derivative order, the semiclassical equations~\eqref{Eq:SemiEqs} can be solved in a full, self-consistent manner. Since the RSET is a function of the spacetime metric and its derivatives, self-consistent solutions to the semiclassical equations will be those in which classical spacetime configurations are everywhere corrected by the backreaction of quantum vacuum polarisation.

Treating semiclassical gravity as a modified theory of gravity generates situations in which semiclassical corrections overcome their $\order{\hbar}$ suppression. 
When there is not classical matter this, for example, can lead to geometries that cannot be smoothly deformed into their general relativistic counterparts in the $\hbar\to0$ limit.
Notwithstanding the absence of a unique prescription to obtain analytic RSET approximations, one advantage of the semiclassical approach is that self-consistent solutions exhibit robust properties that appear to be independent of the RSET approximation adopted. 
Before passing to the description of the solutions found, let us make two remarks regarding the selection of Boulware state and the surpassing of the Buchdahl limit.

\subsubsection{The importance of the Boulware vacuum}

Semiclassical gravity discloses its non-perturbative phenomenology at event horizons. In static situations, the natural vacuum state for the field is the Boulware vacuum~\cite{Boulware1974}, which reduces to the Minkowski vacuum at radial infinity, a characteristic that is consistent with the asymptotic flatness of spacetime. This modes defining the state are manifestly singular at the event horizon, such behaviour spreading to the RSET. In fact, the RSET has a physical divergence at $r=r_{\textrm H}$ if the energy density measured by a freely-falling observer diverges there~\cite{Loranzetal1995}. Let us illustrate this divergence in the Polyakov RSET. For a metric adopting the form
\begin{eqnarray}
    f(r)=h(r)^{-1}=\frac{r-r_{\rm H}}{r_{\rm H}}+\order{\frac{r-r_{\rm H}}{r_{\rm H}}}^{2}
\end{eqnarray}
near the event horizon, then the quantity 
\begin{eqnarray}\label{Eq:FreeFallDens}
    \mathcal{E}=\frac{\langle\hat{T}^{r}_{r}\rangle^{\rm P}-\langle\hat{T}^{t}_{t}\rangle^{\rm P}}{f}\propto -\frac{l_{\textrm P}^{2}}{r_{\textrm H}^{2}\left(r-r_{\textrm H}\right)^{2}}+\mathcal{O}\left(\frac{r-r_{\textrm H}}{r_{\textrm H}}\right)^{0},
\end{eqnarray}
is infinite (here $l_{\rm P}=1/\sqrt{12\pi}$). A similar divergence is found for the AHS-RSET~\cite{Arrecheaetal2023}. 

As a consequence of its divergent behaviour at the event horizon,\footnote{This divergence is rooted to the choice of plane-wave mode solutions to Eq.~\eqref{Eq:WaveEq2D}. The null Eddington-Finkelstein coordinates diverge at even horizons making the modes to acquire infinite frequencies.} the Boulware vacuum is commonly dismissed as plainly non-physical, deemed as the natural state only for horizonless stellar configurations instead. This argument is based on the assumption that the background spacetime is unaffected by vacuum polarisation. At the level of the semiclassical equations, when the RSET is allowed to backreact on the spacetime, the Boulware vacuum is a perfectly self-consistent vacuum state, since its characteristic divergence gets absorbed by the background spacetime. As a consequence, the event horizon gets destroyed by the backreaction of vacuum polarisation. When there is no classical matter, this carries along additional pathologies that are absent in classical vacuum solutions, such as the presence of curvature singularities that are not concealed by event horizons. This is avoided in the presence of classical matter, which makes the geometry akin to stellar configurations.

At this stage, it is interesting to advance an intriguing difference between GR and its semiclassical counterpart~\cite{Arrecheaetal2021b}. 
In GR the eternal Schwarzschild (or Kerr) vacuum solution captures all the relevant aspects of the more realistic situation in which a BH is formed from the collapse of a previous stellar configuration. This can be taken as suggesting that 
the behaviour of matter is not that relevant for analysing the end point of gravitational collapse (beyond imposing that matter does not violate energy conditions). On the contrary, in semiclassical GR vacuum eternal (or static) solutions exhibit more clearly their pathological nature. As we will see, in order to obtain physically reasonable configurations, some classical matter needs to be included in the spacetime. It is this classical material that seeds how vacuum polarisation is in turn excited.

\subsubsection{Surpassing the Buchdahl limit}

We follow this line of thought to its ultimate consequences, assuming the existence of an additional material modelled as a classical SET describing an isotropic perfect fluid of constant density in equilibrium. Through this simple assumption, a window opens towards the possibility that the (on average repulsive) effects of vacuum polarisation generate new configurations in equilibrium that are more compact than those allowed by classical general relativity. We define the compactness function as
\begin{eqnarray}
    C(r)\equiv\frac{2m(r)}{r}= 1-h(r)^{-1},
\end{eqnarray}
where $m(r)$ is the Misner-Sharp mass~\cite{MisnerSharp1964}. The value $C(r)=1$ denotes the compactness of the BH event horizon and $C(r)=8/9$ denotes the largest surface compactness  attainable by hydrostatic equilibrium configurations in general relativity, or Buchdahl limit~\cite{Buchdahl1959, UrbanoVeermae2018}. The Buchdahl compactness bound applies to stars satisfying the following: (i) the star has a Schwarzschild exterior, (ii) internal pressures in the angular directions do not surpass the pressure in the radial direction, and (iii) a density profile that is non-increasing outwards. Self-consistent semiclassical gravity has the potentiality to violate all three possibilities: the exterior spacetime is no longer Schwarzschild, RSETs are anisotropic by construction, and they have negative energy densities that can revert the tendency of the total density to be non-increasing outwards. In consequence, this theory stands out as a promising place in which to seek for new stages of stellar equilibrium that can solve the pathologies posed by vacuum solutions.

By seeking for RSET approximations that are adapted to describe stellar structures, it is possible to find families of RSETs whose backreaction effects support stars that overcome the Buchdahl limit. Once the Buchdahl limit is surpassed, these stars can have a surface lying extremely close to their gravitational radius (both surfaces being separated just by few Planck lengths). Their large interior redshifts makes them easily mistaken for BHs through electromagnetic observations~\cite{Carballo-Rubioetal2018}. Nonetheless, the presence of a surface inside their photon sphere could produce distinct gravitational-wave echoes~\cite{CardosoPani2019}. This way we evidence that, by considering a more physical vacuum than the one from general relativity, together with the simplest material contents, semiclassical gravity allows for the existence of ultracompact alternatives to BHs. 

These exotic compact objects, that we denoted as relativistic semiclassical stars, are realised in two ways: through exploring families of RP-RSETs where the regulator function $F(r)$ is distorted within some central stellar core, and through order-reduced versions of the AHS-RSET. 
Our philosophy in here is not to argue for a particular approximation scheme as the best one; it is more to put all the possibilities on the table to see what they can offer. Our point of view concerning semiclassical theories is more heuristic and closer to the phenomenological philosophy underneath modified theories of gravity: motivating a possible form for some modifications of general relativity and then analysing the new equations without caring how these equations might show up hierarchically from an even deeper description of spacetime. The existence of common features in the solutions to semiclassical equations sourced by unrelated RSETs evidences the robustness of semiclassical analyses. 

\subsection{Solutions with no classical matter}

We now turn towards deriving the semiclassical counterpart to the Schwarzschild BH solution. By semiclassical counterpart, we refer to the solution that incorporates the effects of vacuum polarisation in a self-consistent way through the backreaction of the RSET. We will use two qualitative approximations to the RSET: the Regularised Polyakov approximation and an Order Reduced AHS approximation. 

\subsubsection{The Regularised Polyakov approximation}

For the metric~\eqref{Eq:4DMetric}, the $tt$ and $rr$ components of the semiclassical Einstein equations in vacuum are, respectively,
\begin{align}\label{Eq:VacuumEqs}
     \frac{h(1-h)-rh'}{h^{2}r^{2}}=
    &
    ~8\pi\hbar\langle \hat{T}^{t}_{t}\rangle^{\rm P},\nonumber\\
    \frac{rf'f-fh}{fhr^{2}}=
    &
    ~8\pi\hbar\langle \hat{T}^{r}_{r}\rangle^{\rm P},
\end{align}
where the RSET is described by the Regularised Polyakov approximation~\eqref{Eq:PolyakovRSET} with 
\begin{eqnarray}\label{Eq:CutoffReg}
    F(r)=1/\left[4\pi\left(r^{2}+\alpha l_{\rm P}^{2}\right)\right],~\alpha>1.
\end{eqnarray}
This simple choice of $F(r)$ acts as a cutoff to the magnitude of the Polyakov RSET, which becomes finite on regular spacetimes. 

\begin{figure}
    \centering
    \includegraphics[width=0.6\columnwidth]{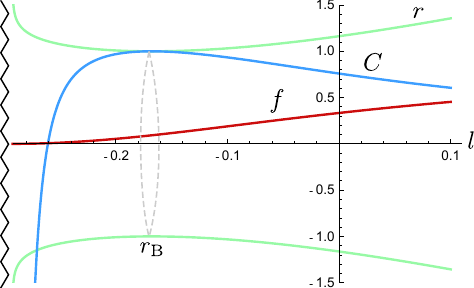}
    \caption{Numerical plot of the semiclassical counterpart of the Schwarzschild vacuum geometry. The horizontal axis is the proper coordinate $l$ while the above and below curves in green represent the radial coordinate $r$. The behavior of the redshift function, in red, and the compactness, in blue, are shown. The right side of the wormhole is asymptotically flat whereas the other is asymptotically singular. Both regions are joined by a minimal surface of radius $r=r_{\rm B}$. We have chosen $M=0.1$ and $\alpha=1.01$ for illustrative purposes. Geometrical characteristics are identical for larger ADM masses.}\label{Fig:Wormh}
\end{figure}

Equations~\eqref{Eq:VacuumEqs} can be integrated as a boundary value problem from radial infinity (in practice, a distant referential radius) assuming the metric takes the asymptotic form
\begin{eqnarray}\label{Eq:SchwMetric}
    f(r)=h(r)^{-1}=1-\frac{2M}{r},\quad M>0,
\end{eqnarray}
which is consistent with the way the RSET components~\eqref{Eq:PolyakovRSET} decay at infinity in the Boulware vacuum~\cite{Arrecheaetal2020}. Due to the presence of an additional source in the right-hand side of the semiclassical equations, the metric no longer obeys $f(r)=h(r)^{-1}$. As the semiclassical equations are integrated inwards, the spacetime geometry progressively deviates from the Schwarzschild solution. This deviation amounts to a redshift function $f(r)$ modified with respect to its Schwarzschild form~\eqref{Eq:SchwMetric} and a Misner-Sharp mass $m(r)$ that acquires a dependence on the radial coordinate, as if the whole spacetime was surrounded by an inhomogeneous cloud of negative mass. 
Notice however that at any macroscopic distance from the gravitational radius this modification is absolutely negligible (it is of order $\hbar$) leaving unchanged any gravitational test related to the Schwarzschild exterior metric.
The magnitude of the RSET increases inwards until a special surface $r_{\rm B}>2M$ where quantum corrections become non-perturbative is encountered. At $r_{\rm B}$, $h(r_{\rm B})\to\infty$ and we find a coordinate singularity that corresponds to a minimal surface for $r$. This is demonstrated by adopting a change to the proper radial coordinate $l$,
\begin{eqnarray}
    \frac{dr}{dl}=\pm\frac{1}{\sqrt{h}},
\end{eqnarray}
where the $\pm$ signs denote the two branches of the radial coordinate at each side of the minimal surface. Assuming the following behaviours for the metric functions
\begin{equation}
	\begin{split}
		f(l)&=f_{\rm B}+f_{1}\left(l-l_{\rm B}\right)+\mathcal{O}\left(l-l_{\rm B}\right)^{2},\\ r(l)&=r_{\rm B}+r_{1}\left(l-l_{\rm B}\right)+r_{2}\left(l-l_{\rm B}\right)^{2}+\mathcal{O}\left(l-l_{\rm B}\right)^{3},
	\end{split}
\end{equation}
and replacing them in the semiclassical equations we find 
\begin{align}\label{Eq:WormhMetric}
    f_{1}=
    \frac{2f_{\rm B}\sqrt{r_{\rm B}^{2}+\alpha l_{\rm P}^{2}}}{l_{\rm P}r_{\rm B}},\quad
    r_{1}=
    0,\quad r_{2}=\frac{\left(r_{\rm B}^{2}+\alpha l_{\rm P}^{2}\right)^{2}+\alpha l_{\rm P}^{4}}{2r_{\rm B}\left[r_{\rm B}^{2}+\left(\alpha-1\right)l_{\rm P}^{2}\right]\left(r_{\rm B}^{2}+\alpha l_{\rm P}^{2}\right)},
\end{align}
where $f_{\rm B}$ and $r_{\rm B}$ are positive constants baring a non-analytic relation to the ADM mass $M$ and the parameter $\alpha$. The functions~\eqref{Eq:WormhMetric} make the metric explicitly regular at $l=l_{\rm B}$. The radial function $r(l)$ is symmetric around the minimal surface $r_{\rm B}$, while the redshift function is positive and asymmetric, showing there is a wormhole neck that connects the asymptotically flat region of the spacetime with a new region of different characteristics. Figure~\ref{Fig:Wormh} shows these metric functions in terms of $l$ for an example integration of the semiclassical equations.

Below the wormhole neck, vacuum polarisation enters into a runaway regime that makes the metric approach a null singularity at infinite $r$, but finite $l$. Through an asymptotic analysis of the semiclassical equations~\cite{Arrecheaetal2020}, the form of the metric nearing the singularity is found to be, in Schwarzschild coordinates,
\begin{equation}\label{Eq:MetricNullSing}
	\begin{split}
		ds^{2}&\simeq\left(\frac{r}{l_{\rm P}}\right)^{1-4\alpha}e^{-\frac{2r^{2}}{l_{\rm P}^{2}}}\cdot \\ & \cdot \left\{-a_{0}\left(1-\frac{l_{\rm P}^{2}}{8 r^{2}}\right)dt^{2}+\frac{2\chi_{0} r^{2}}{l_{\rm P}^{2}}\left[1-\frac{(9-32\alpha)l_{\rm P}^{2}}{8r^{2}}\right]dr^{2}\right\}+r^{2}d\Omega^{2},
	\end{split}
\end{equation}
where $a_{0}$ and $\chi_{0}$ are dimensionless positive constants. The vanishing of the conformal factor as $r\to\infty$ manifests the null character of this singularity. The Ricci scalar, defined as
\begin{eqnarray}\label{Eq:RicciScalar}
    \mathcal{R}=\frac{2}{r^{2}}\left(1-\frac{1}{h}\right)+\frac{2}{hr}\left(\frac{h'}{h}-\frac{f'}{f}+\frac{rf'h'}{4fh}\right)+\frac{1}{2h}\left[\left(\frac{f'}{f}\right)^{2}-2\frac{f''}{f}\right], 
\end{eqnarray}
becomes negatively divergent at the singularity, i.e.,
\begin{eqnarray}
\mathcal{R}\simeq-\frac{e^{2r^{2}/l_{\textrm P}^{2}}(2\alpha-1)}{l_{\textrm P}^{2}\chi_{0}}\left(\frac{r}{l_{\textrm P}}\right)^{-5+4\alpha}.
\end{eqnarray}
Furthermore, this singular region is located at a finite proper distance $l_{\rm S} < l_{\rm B}$ from the throat, as shown by integrating the quantity
\begin{eqnarray}
     \left(\frac{dl}{dr}\right)^{2}=2\chi_{0}\left(r/l_{\rm P}\right)^{3-4\alpha}e^{-\frac{2r^{2}}{l_{\rm P}^{2}}}\left[1-\frac{(9-32\alpha)l_{\rm P}^{2}}{8r^{2}}\right].
\label{Eq:properdistance}
\end{eqnarray}

\begin{figure}
    \centering
    \includegraphics[width=0.8\columnwidth]{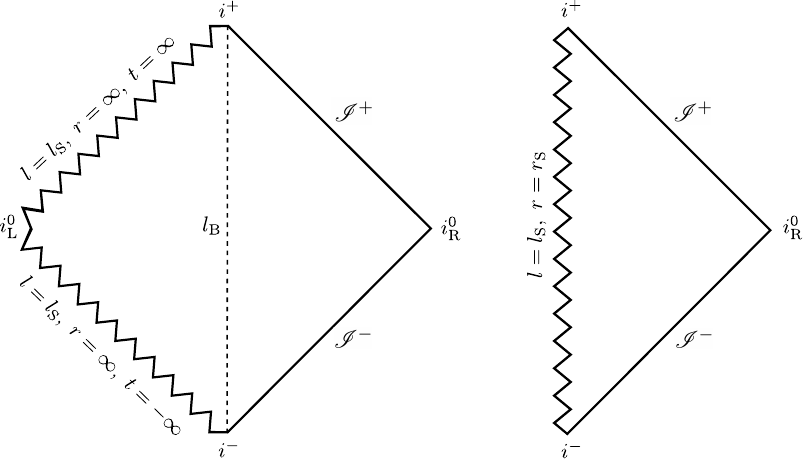}    
    \caption{Left panel: Penrose diagram corresponding to the singular wormhole solution for the RP-RSET. The dashed lines denote the location of the wormhole neck. To their right, the asymptotically flat portion of spacetime is depicted alongside its asymptotic regions. The left hand side of the diagram shows the internal past and future null singularities, which are located at finite proper distance from the neck $l_{\textrm S}-l_{\textrm B}$. The point $i^{0}_{\textsc{l}}$ is singular as well, and is reached in finite proper time by spacelike geodesics. Right panel: Penrose diagram associated with the vacuum solution for the OR-RSET.
    In this case, the singularity is timelike and constitutes a naked singularity. While differences in the modelling of the semiclassical source result in singularities of different sorts, both models agree on the absence of event horizons.}\label{Fig:Penrose}
\end{figure}

Near asymptotically flat regions, far from any source of gravity, semiclassical corrections amount to extremely weak, thus perturbative, corrections. As the surface $r=2M$ is approached, however, vacuum polarisation builds up and destroys the event horizon, generating instead a wormhole neck. While, as shown below, the specific features of the region of non-perturbative semiclassical corrections depend on the particular modelling of the RSET (see Fig.~\ref{Fig:Penrose}), the replacement of the event horizon by a singularity appears as a robust characteristic of the RSET not depending on the approximation, as it is a consequence of evaluating the RSET in the Boulware state. Similar characteristics are displayed by the semiclassical counterpart to the Reissner-Nordström sub-extremal BH~\cite{Arrecheaetal2021}.

The regularisation scheme we have adopted for the Polyakov RSET [see Eq.~\eqref{Eq:CutoffReg}] amounts to a cutoff to the magnitude of its components, whose strength is modulated by $\alpha$. Increasing $\alpha$ brings the singularity at $l_{\rm S}$ closer to the wormhole neck $l_{\rm B}$. Spacetime regions near $r=0$, while unexplored in vacuum solutions, are present in stellar configurations. We will return to exploring more elaborate regularisation schemes for the Polyakov RSET later. 

The results here presented are consistent with previous works~\cite{Fabbrietal2006,Berthiereetal2017,HoMatsuo2017}. A clear extension of this work is to consider the backreaction effects of an RSET approximation that does not rely on dimensional reduction. The Order Reduced version of the AHS-RSET stands on equal footing with the RP-RSET, in the sense that it is a quantity covariantly conserved, analytic, without higher-derivative terms, and well-defined at $r=0$. We now briefly sketch the derivation of this RSET approximation and the characteristics of the corresponding vacuum solutions in the minimally coupled case.  

\subsubsection{The Order Reduced AHS-RSET}

We start with the $tt$ and $rr$ components of the vacuum semiclassical equations~\eqref{Eq:SemiEqs}, now sourced by the AHS-RSET,
\begin{align}\label{Eq:SemiComps}
     \frac{h(1-h)-rh'}{h^{2}r^{2}}=
    &
    ~8\pi\hbar\langle \hat{T}^{t}_{t}\rangle^{\text{AHS}},\nonumber\\
    \frac{rf'f-fh}{fhr^{2}}=
    &
    ~8\pi\hbar\langle \hat{T}^{r}_{r}\rangle^{\text{AHS}},
\end{align}
where the right-hand side contains higher-derivative terms. The concrete and lengthy form of the AHS-RSET, which is not very illustrative, can be seen in~\cite{Andersonetal1995}. To obtain a set of equations of the same derivative order as the classical ones, we subject the AHS-RSET to a perturbative reduction of order. 
The first step in this procedure consists in neglecting terms $\order{\hbar}$ in Eq.~\eqref{Eq:SemiComps}, leading to
\begin{align}\label{Eq:hbarExp}
    \frac{h(1-h)-rh'}{h^{2}r^{2}}=
    &
    ~\order{\hbar},\nonumber\\
    \frac{rf'+f-fh}{fhr^{2}}=
    &
    ~\order{\hbar}.
\end{align}
These expressions can be differentiated consecutively to derive recursion relations between $f,~h$, and their higher-order derivatives $\{f^{(n)}\}_{n=1}^\infty$ and $\{h^{(n)}\}_{n=1}^\infty$. For $h$, said relations are obtained by solving the $tt$ equation directly, which can then be used to derive the $f$ relations from the $rr$ equation. The resulting relations are
\begin{align}\label{Eq:OrderRels}
    h^{(n)}=
    &
    \left(-1\right)^{n}\frac{n!h^{n}}{r^{n}}\left(h-1\right)+\order{\hbar},\nonumber\\
    f^{(n)}=
    &
    \left(-1\right)^{n+1}\frac{n!f}{r^{n}}\left(h-1\right)+\order{\hbar}.
\end{align}

Relations \eqref{Eq:OrderRels} are now inserted in the AHS-RSET components $\langle\hat{T}^{t}_{t}\rangle^{\text{AHS}}$ and $\langle\hat{T}^{r}_{r}\rangle^{\text{AHS}}$
until they only depend on $f$ and $h$.
After a lengthy but straightforward calculation using symbolic computation software, we arrive at
\begin{align}\label{Eq:VOR}
    16\pi^{2}\langle\hat{T}^{t}_{t}\rangle^{\textrm {OR}}=
    &
    \frac{\left(h-1\right)^{2}\left(h^{2}+6h+33\right)}{480h^{2}r^{4}}
    -\left(\xi -\frac{1}{6}\right)\frac{\left(h-1\right)^{2}\left(h^{2}+2h+5\right)}{8h^{2}r^{4}},\nonumber\\
    16\pi^{2}\langle\hat{T}^{r}_{r}\rangle^{\textrm {OR}}=
    &
    -\frac{\left(h-1\right)^{2}\left(h^{2}+6h-15\right)}{1440h^{2}r^{4}}
    +\left(\xi-\frac{1}{6}\right)\frac{\left(h-1\right)^{2}\left(h+3\right)^{2}}{24h^{2}r^{4}},
\end{align}
where the suffix OR stands for Order-Reduced.
Using the expressions for $\langle\hat{T}^{t}_{t}\rangle^{\textrm{OR}}$ and $\langle\hat{T}^{r}_{r}\rangle^{\textrm{OR}}$  we can deduce, through the conservation relation~\eqref{Eq:ConsRSET}, the angular components necessary to force its divergence to vanish. Specifically, we find 
\begin{align}\label{Eq:beyond:Tthth}
16\pi^{2} & \langle\hat{T}^{\theta}_{\theta}\rangle^{\textrm{OR}}=
16\pi^{2}\langle\hat{T}^{\varphi}_{\varphi}\rangle^{\textrm{OR}}= \nonumber \\
&
-\frac{h-1}{1440fh^{2}r^{4}}\left\{f\left[h^{3}\left(r-1\right)+h^{2}\left(3r-5\right)+3h\left(r+7\right)-15\left(r+1\right)\right]\right. \nonumber\\
&
~~~~~~~~~~~~~~~~~~~~~~~~~~~\left.+f'\left(h-1\right)\left(h^{2}+6h+21\right)\right\} \nonumber\\
&
+\left(\xi-\frac{1}{6}\right)\frac{h-1}{24fh^{2}r^{4}}\left\{f\left(h+3\right)\left[h^{2}\left(r-1\right)-2h+3\left(r+1\right)\right]\right.
\nonumber\\
&
~~~~~~~~~~~~~~~~~~~~~~~~~~~\left.+rf'\left(h-1\right)\left(h^{2}+3h+6\right)\right\}.
\end{align}
In this way, we construct a new approximate RSET (the OR-RSET) valid for backreaction studies in vacuum spacetimes and which satisfies all Wald's axioms~\cite{Wald1977}. We can now assume the scalar field is minimally coupled to gravity ($\xi=0$) to analyse a material content equivalent to the one described by the Regularised Polyakov approximation. 

Regarding the semiclassical counterpart to the Schwarzschild spacetime in this prescription, it is possible to construct an analytical argument~\cite{Arrecheaetal2023} proving that, when the order-reduced semiclassical equations are integrated from the asymptotically flat region inwards, the solution inevitably encounters a naked, timelike singularity at $r=r_{\rm D}>r_{\rm H}$. Roughly sketched, this argument is based on the fact that the $tt$ equation enforces the $h$ function to be monotonically increasing inwards. By ruling out any other possibility, we find it must diverge at $r=r_{\rm D}$. This implies, through the $rr$ equation, that $f$ has a global minimum at some radius $r>r_{\rm D}$, diverging positively at the singularity. Near the singular region, the metric behaves as
\begin{eqnarray}\label{Eq:MetricSing}
    f\simeq\left(\frac{f_{\rm D}}{r-r_{\rm D}}\right)^{1/9},\quad h\simeq\left(\frac{h_{\rm D}}{r-r_{\rm D}}\right)^{1/3},
\end{eqnarray}
where $f_{\rm D},~h_{\rm D}$ are positive constants bearing a relation to the ADM mass $M$ and $r_{\textrm D}$, not relevant here. 
The behaviours~\eqref{Eq:MetricSing} cause the curvature scalar~\eqref{Eq:RicciScalar} to diverge as
\begin{eqnarray}
    \mathcal{R}\simeq -\chi_{1}\left(r-r_{\rm D}\right)^{-5/3},\quad \chi_{1}>0.
\end{eqnarray}
This divergence is slower than the one predicted by the Polyakov approximation, but it is reached at finite $r$ and finite affine distance for all geodesic observers, giving the singularity a timelike character. The Penrose diagram associated to this solution is represented in Fig.~\ref{Fig:Penrose}. 

Through two unrelated methods for obtaining proper RSET approximations in static situations for massless, minimally coupled fields, we have shown that the semiclassical counterparts to the Schwarzschild spacetime exhibit pathologies that make them inadequate to describe the late time outcome of gravitational collapse. Either they present null curvature singularities and a change in the topology of the spacetime, as for the RP-RSET, or they exhibit a naked curvature singularity, as in the OR-RSET. Note that this geometry is obtained for as long as $\xi<11/60$.

There are two possibilities for making sense of semiclassical solutions in light of these results: Either we consider horizons are dynamic and undergo an evaporation process or, if we insist on exploring static possibilities within this theory, introducing classical matter becomes essential.

\subsection{Stellar equilibrium}

After considering purely vacuum solutions in semiclassical gravity, i.e., geometries self-consistent with the cloud of vacuum polarisation that they generate, we turn to the analysis of stellar configurations in hydrostatic equilibrium. Adding classical matter makes the analysis of the semiclassical equations considerably more involved, since the classical and quantum SETs affect each other through the geometry of spacetime. In the aim to simplify this analysis, we model the stellar interior via
the stress-energy tensor (SET) of an isotropic perfect fluid
\begin{eqnarray}\label{Eq:ClassicalSET}
    T^{\mu}_{\nu}=\left(\rho+p\right)u^{\mu}u_{\nu}+p\delta^{\mu}_{\nu},
\end{eqnarray}
with $p$ and $\rho$ denoting the pressure and energy density measured by an observer comoving with the static fluid with $4$-velocity $u^{\mu}$. Covariant conservation of the SET~\eqref{Eq:ClassicalSET} translates into the relation
\begin{eqnarray}\label{Eq:Cont}
    \nabla_{\mu}T^{\mu}_{r}=p'+\frac{f'}{2f}\left(\rho+p\right)=0.
\end{eqnarray}
An equation of state for the classical fluid needs to be specified. We assume a simple uniform-energy-density fluid
\begin{eqnarray}\label{Eq:EoS}
    \rho(r)\equiv\rho=\text{const},
\end{eqnarray}
so that we are seeking for the semiclassical counterparts of Schwarzschild's stellar solutions~\cite{Schwarzschild1916}. In standard GR, the assumption of an interior with uniform classical density saturates the hypotheses behind the Buchdahl theorem: it produces the most compact regular stars, with $C(R)=8/9$. In our analyses, any surpassing of the Buchdahl compactness bound will be caused by the semiclassical corrections we have incorporated through the RSET. 

We start integrating the semiclassical equations~\eqref{Eq:SemiEqs} from the asymptotically flat region inwards in absence of any classical SET and with positive ADM mass. This vacuum geometry will correspond to either of the solutions found in the previous section, depending whether we consider the RP-RSET or the OR-RSET. Then, at a particular $R$ we continue the integration but now in the presence of the classical SET, that is, we consider the equations
\begin{align}\label{Eq:SemiEqsMatter}
    \frac{h(1-h)-rh'}{h^{2}r^{2}}=
    &
    8\pi\left(-\rho+\hbar\langle\hat{T}^{t}_{t}\rangle\right),\nonumber\\
    \frac{rf'+f-fh}{fhr^{2}}=
    &
    8\pi\left(p+\hbar\langle\hat{T}^{t}_{t}\rangle\right).
\end{align}
The RSET will, in this interior region, be modelled in two ways that are natural generalisations of the RSETs we considered in the vacuum case.

Since the standard Polyakov approximation is not well-defined at $r=0$, some regularisation of this approximation is mandatory to construct stellar equilibrium configurations, even in those situations where the semiclassical corrections received are small and perturbative. Therefore, the behaviour of the Polyakov approximation at $r=0$ needs to be specified by hand through suitable choices of $F(r)$. To this aim, we consider modified $F(r)$ functions that distort the Polyakov approximation in some internal region of the star $r\leq r_{\text{core}}<R$ (the stellar core hereafter). Through minimal assumptions on the geometrical characteristics of this core, we find, among the space of modified $F(r)$ functions, whole families compatible with regular stars that surpass the Buchdahl limit (see next subsection).

Then, to add cumulative evidence to this result and erase all doubt of it being an ad hoc procedure, we derive the equivalent to the previous OR-RSET but now in the presence of a classical constant density fluid. Sourced by this RSET, we find regular stars that surpass the Buchdahl limit with remarkably similar properties to those obtained through the RP-RSET (subsection \ref{Subsubsec:order-reduced}). 

Exploring the space of semiclassical stellar solutions is an exhaustive task than involves both regular and singular solutions. For didactic purposes, here we just outline the derivation of the corresponding RSETs supporting semiclassical stars that surpass the Buchdahl limit. In between, we discuss the main physical properties of these new objects and the implications stemming from them. 
We encourage the reader to consult the references~\cite{Arrecheaetal2021c,Arrecheaetal2022,Arrechea2023} for details. 

\subsubsection{Stellar equilibrium in a regularised Polyakov approximation}

We now consider the semiclassical equations~\eqref{Eq:SemiEqsMatter} in presence of classical matter which occupies a finite portion of spacetime inside a sphere of radius $R$. The geometry exterior to this sphere is the Schwarzschild counterpart in the Regularised Polyakov approximation. For simplicity, let us place the star surface at the side of the wormhole that connects with the asymptotically flat region, although it is also possible to place it at the neck and even slightly inside. 

The integration from the surface inwards takes $C(R)$ and $\rho$ as parameters. The regulator function $F(r)$ is first taken, for simplicity, to be of the form~\eqref{Eq:CutoffReg}.
Sweeping through the parameter space we find for $C(R)<8/9$, regular stars perturbatively close to their classical counterparts. This is to be expected from their small interior curvatures that generate a negligible polarisation of the vacuum. Increasing the surface compactness to the regime $1>C(R)\geq8/9$, we stop finding regular stars. Nonetheless, it is always possible to observe a finite set of $\rho$ values for which stars contains a surface where $C(r_{0})=0$ at $0<r_{0}\ll R$. We denote these solutions as ``quasi-regular" stars, given that a vanishing compactness at $r=0$ is a requirement for spacetime regularity\footnote{Strictly, regularity conditions impose that compactness must vanish at least as $C\simeq C_{0}r^{2}+\order{r^{3}}$ and redshift must go as \mbox{$f\simeq f_{0}+f_{1}r^{2}+\order{r^{3}}$}. Conditions on the pressure are then derived from Eq.~\eqref{Eq:Cont}.}. Since 
$r_{0}$ can be arbitrarily small, this is taken as an indication that semiclassical effects bring super Buchdahl stars closer to attaining regularity than what classical matter can do by itself (quasi-regular solutions are not found in the classical theory in the compactness range $1>C(R)\geq8/9$).

From a physical perspective, adding a small, positive $\rho$ to the spacetime at $R>r_{\rm B}$ effectively shrinks the size of the wormhole neck slightly, which now appears inside the region covered by matter. As $\rho$ increases, this wormhole neck shrinks further, eventually disappearing for some critical density $\rho_{\rm c}$. Quasi-regular stars are those displaying a wormhole neck extremely close to the centre of spherical symmetry. These configurations are realised due to a fine balance between classical and quantum contributions that eventually breaks down in the innermost regions of the star, precisely where the Polyakov approximation is less reliable. 

The regularisation adopted in~\eqref{Eq:CutoffReg} is just the simplest among plenty of possibilities, some of which might even approximate the behaviour of an exact RSET in these regions. In what follows, we search for more elaborate regularisation schemes for the RP-RSET. These regularisations are characterised by the following: for any star belonging to or nearby the quasi-regular family, we select a spherical surface of radius $r_{\text{core}}\in(r_{0},R)$ and, in the range $r\in[0,r_{\text{core}})$, we assume minimal deformations of the regulator function~\eqref{Eq:CutoffReg} that are compatible with the complete absence of curvature singularities.

To find the form of these deformations of the $F$ function, we follow a reverse-engineering logic that consists in proposing an ansatz for a regular geometry in the range $r\in[0,r_{\text{core}})$ and then obtaining the regulator $F$ that sources the geometry via the RP-RSET, in case there is any compatible with regularity. For this purpose, we derive an expression for $h$ from the $rr$ component of the semiclassical equations~\eqref{Eq:SemiEqsMatter} and replace it in the $tt$ component. Through the conservation relation of the classical SET~\eqref{Eq:Cont}, we arrive at the expression
\begin{align}
    p''=
        \mathcal{D}\left[\mathcal{A}_{0}+\mathcal{A}_{1}\left(p'\right)+\mathcal{A}_{2}\left(p'\right)^{2}+\mathcal{A}_{3}\left(p'\right)^{3}\right],
    \label{Eq:PresSemi}
\end{align}
where 
\begin{align}
    \mathcal{A}_{0}=
    &
    -8\pi r\left(\rho+p\right)^{3}\left(\rho+3p\right),\nonumber\\
    \mathcal{A}_{1}=
    &
    ~4\left(\rho+3p\right)^{2}\left[6\pi r^{2}\left(\rho+p\right)+16\pi^{2} F l_{\textrm P}^{2}r^{2}p -1\right],\nonumber\\
    \mathcal{A}_{2}=
    &
    -r\left(\rho+p\right)\left[16\pi r^{2}\left(\rho-2p\right)-4\pi l_{\textrm P}^{2}\left(2F+rF'\right)\right.\nonumber \\
    &\quad\left.+32\pi^{2} Fl_{\textrm P}^{2} r^{2}\left(\rho+5p\right)-32\pi^{2} F'l_{\textrm P}^{2}r^{3}p-6\right],\nonumber\\
    \mathcal{A}_{3}=
    &
    ~F l_{\textrm P}^{2}r^{2}\left[8\pi r^{2}\left(\rho-p\right)-4\pi l_{\textrm P}^{2}\left(2F-rF'\right)-32\pi^{2}F'l_{\textrm P}^{2}r^{3}p-2\right],\nonumber\\
    \mathcal{D}=
    &
    ~2r\left(1-4\pi l_{\textrm P}^{2}F\right)\left(\rho+p\right)^{2}\left(1+8\pi r^{2}p\right).
\end{align}
Now we make use of this differential equation to find the geometry and the RP-RSET inside the core. By imposing an ansatz for the pressure and its derivatives in the region $r\in\left[0,r_{\text{core}}\right)$, Eq.~\eqref{Eq:PresSemi} becomes a first order differential equation for $F$ which, upon solving, determines the entire geometry inside the core. Naturally, if the resulting $F$ and $F'$ are everywhere finite within the core, $h(r)$ is regular from~\eqref{Eq:SemiEqsMatter}, and hence the complete spacetime metric.

We consider a pressure profile for the core [whose classical energy density is constant, recall Eq.~\eqref{Eq:EoS}] that is everywhere finite and has a global maximum at $r=0$. These are the minimal conditions necessary for regularity, given that we are just extending the behaviour of the pressure function in quasi-regular stars (which grows monotonically from the surface inwards) in the simplest way, that is, so that it reaches a global maximum at $r=0$.
At $r_{\text{core}}$, continuity of the metric enforces pressure to be continuous up to its second derivative.
The simplest analytic function that satisfies these conditions is the fifth-order polynomial
\begin{eqnarray}\label{Eq:Polynomialfit}
p= p_{0}+p_{0}''r^{2}/2+c_{0} r^3+c_{1}r^4+c_{2}r^5,
\end{eqnarray}
where the pressure at the origin $p_{0}$ and its second derivative $p_{0}''$ are positive and negative constants, respectively. Determining the coefficients $\left\{c_{i}\right\}_{i=0}^{2}$ is straightforward given the aforementioned conditions of regularity and continuity. 

The last step consists in taking a fixed numerical solution for the bulk region $r\in[r_{\text{core}},R)$. This amounts to choosing a particular pressure profile, which we typically consider to be the critical solution $\rho=\rho_{\rm c}$, but we can also select a nearby solution in the space of parameters. Given a core size $r_{\text{core}}$, the pressure function inside the core is determined upon fixing the two remaining free parameters $\{p_{0},p_{0}''\}$ by hand.

We explore numerically a wide range of values of the parameters $\{p_{0},p_{0}''\}$ that select a particular pressure profile in~\eqref{Eq:Polynomialfit} and obtain their associated $F(r)$ through~\eqref{Eq:PresSemi}. We find that regular solutions exist for cores of any size $r_{\text{core}}$. Let us emphasise that the existence of regular semiclassical stars that surpass the Buchdahl limit is a remarkably non-trivial result. Of course, the Einstein equations guarantee that any exotic spacetime geometry, like that of a BH mimicker, is generated by a corresponding effective SET that will certainly violate energy conditions. This should not be mistaken with our approach here, as we are imposing that the pressure inside the core obeys a polynomial form and asking whether this is compatible with a stress-energy tensor that can be divided in two parts: a classical uniform density perfect fluid and a RP-RSET. As it is not guaranteed that a prescribed geometry will be compatible with the RP-RSET, it might have happened that no $F$ existed for any regular pressure ansatz. Hence, that this compatibility is realised for the simple polynomial example in Eq.~\eqref{Eq:Polynomialfit} is a strong indication that the Polyakov approximation is able to capture an important fraction of the relevant physics. Furthermore, the existence of these solutions for central cores of any size suggests that semiclassical physics is capable of arranging itself in the way necessary to attain equilibrium in each situation.

\subsubsection{Physical properties of semiclassical stars}

\begin{figure}
    \centering
    \includegraphics[width=\columnwidth]{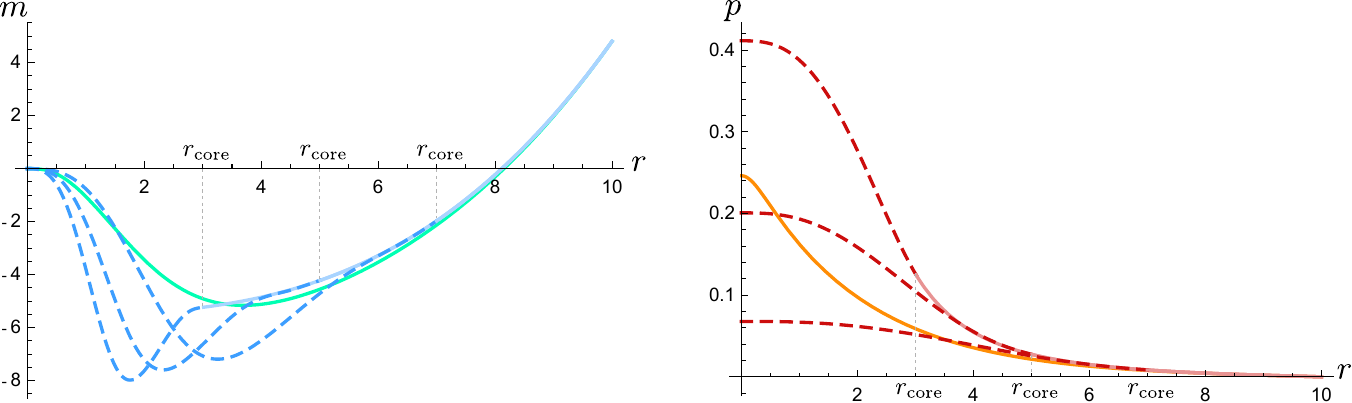}
    \caption{Misner-Sharp mass (left plot) and classical pressure (right plot) of stars with $C(R)=0.96$ and $R=10$. The dashed curves represent solutions obtained through the Regularised Polyakov approximation. We have shown three solutions with $\rho\simeq0.00246$ and different $r_{\text{core}}$ values, below which the $F$ function is distorted. We have also plotted a star obtained in the order-reduced prescription with $\rho\simeq0.00248$ (turquoise and orange lines) that highlight the qualitative agreement obtained through both RSET approximations.}\label{Fig:Star}
\end{figure}

Figure~\ref{Fig:Star} shows the Misner-Sharp mass and classical pressure profiles for semiclassical relativistic stars with different core sizes. Stars with larger cores have less negative Misner-Sharp mass, although these values are distributed throughout a wider portion of the stellar interior. Increasing the size of the core also diminishes the magnitude of the central pressure $p_{0}$.

\begin{figure}
    \centering
    \includegraphics[width=0.6\columnwidth]{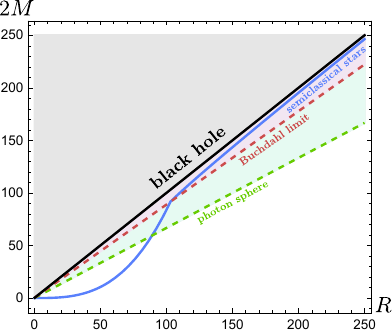}
    \caption{Mass-to-Radius diagram of semiclassical stars in the Regularised Polyakov approximation with $\rho=10^{-5}$ (in Planck units). The black line represents the compactness parameter of BHs, $C(R)=1$, the dashed red line denotes the Buchdahl compactness bound \mbox{$C(R)=8/9$}, and the dashed green line is the minimum compactness of objects that exhibit a photon sphere, $C(R)=1/3$. The blue curve represents semiclassical stars. For those stars surpassing the Buchdahl limit, the total mass $M$ grows approximately linearly with the radius $R$. Each point within the blue curve admits whole families of regulator functions $F$ for which the entire geometry is regular. A qualitatively similar diagram is obtained making use of the MOR-RSET.}\label{Fig:MtoR}
\end{figure}

Semiclassical contributions are suppressed by $l_{\textrm P}$ so, in order for vacuum polarisation to be an appreciable effect, there needs to be some additional scale compensating for this suppression. In the case of semiclassical stars, this scale is precisely the proximity (in compactness) to the Buchdahl limit. When the surface compactness of a classical constant density sphere reaches the value $C(R)=8/9$, spacetime curvature diverges at $r=0$, triggering a subsequent divergence in the RSET components~\cite{ReyesTomaselli2023}. The backreaction of this diverging RSET can then incorporate additional repulsive effects that prevent the pressure from diverging, allowing the existence of regular stars at the Buchdahl limit and beyond. This marks a transition between regimes with negligible and significant quantum corrections, which can be observed in the mass-radius diagram from Fig.~\ref{Fig:MtoR}. In this diagram, we find, in addition to the perturbatively corrected stars with $C(R)<8/9$, whole families of semiclassical stars whose surface compactness ranges from the Buchdahl limit $C(R)=8/9$ to the BH limit $C(R)=1$ itself. The phenomenological implications of this result are promising, as they allow to probe the observational properties of exotic compact objects at nearly unexplored compactness values~\cite{Konoplyaetal2019}.

Looking at the bigger picture, semiclassical stars are the result of a balance between classical and quantum effects. A promising extension of this work is to analyse whether similar solutions exist when using another RSET approximations such as the AHS-RSET, which is four dimensional~\cite{Andersonetal1995} by construction. This RSET is a perfectly well-defined source in stellar spacetimes, with the shortcoming that it exhibits high-order terms in the derivatives of the metric. To conclude this section, we present below a method of reduction of order that transforms the high-derivative terms in the AHS-RSET into low-derivative ones, yielding a novel RSET approximation that is amenable to backreaction studies in stellar spacetimes. Using this approximation, we have found preliminary evidence for the existence of stellar spacetimes with akin characteristics (i.e. surpassing the Buchdahl limit by means of a negative mass interior). 

\subsubsection{Stellar equilibrium in an order-reduced prescription}
\label{Subsubsec:order-reduced}

To construct an order-reduced version of the AHS-RSET for stellar spacetimes, take the $tt$ and $rr$ components of the semiclassical equations~\eqref{Eq:SemiEqsMatter} (now sourced by the AHS-RSET) and neglect terms of $\order{\hbar}$, such that
\begin{align}\label{Eq:beyond:hbarExpMat}
    \frac{h(1-h)-rh'}{h^{2}r^{2}}=
    &
    -8\pi\rho+\order{\hbar},\nonumber\\
    \frac{rf'+f-fh}{fhr^{2}}=
    &
    ~8\pi p+\order{\hbar}.
\end{align}
Consecutively differentiating these expressions and replacing derivatives of $p$ by derivatives of $f$ through Eq.~\eqref{Eq:Cont}, we derive a set of expressions relating derivatives of any order of $f,h$, and $p$, with these functions themselves [these take a more involved form than the simple recursion relations from Eq.~\eqref{Eq:OrderRels}].
Upon replacing said relations in the AHS-RSET, its $\langle\hat{T}^{t}_{t}\rangle$ and $\langle\hat{T}^{r}_{r}\rangle$ components are reduced to expressions without spatial derivatives. Lastly, we obtain low-derivative-order expressions for the $\langle\hat{T}^{\theta}_{\theta}\rangle$ component by imposing covariant conservation through~\eqref{Eq:ConsRSET}, in a manner analogous to the OR-RSET. This leads to the Matter-Order-Reduced RSET (MOR-RSET), a semiclassical source that is analytic, conserved, low-order and well-defined at $r=0$. These characteristics put the MOR-RSET on equal footing with regularised versions of the Polyakov approximation. 
More details on the procedure and full expressions are available in the PhD Thesis~\cite{Arrechea2023}, and will appear in a forthcoming publication.

The AHS- and MOR-RSETs are valid to describe scalar fields with any curvature coupling. Here, we assume a massless, minimally coupled scalar field to allow for a faithful comparison with the RP-RSET. Semiclassical stars that surpass the Buchdahl limit are also found for non-minimally-coupled fields, but this is something that we leave for future studies on the subject.

Sourced by the MOR-RSET,
the semiclassical equations are of the same derivative order as the classical Einstein equations and can be solved as a boundary value problem, assuming the exterior geometry is given by the OR-RSET. By specifying the integration parameters $R$,~$C(R)$ and $\rho$, we perform an exploration of the space of solutions and find regular stars that surpass the Buchdahl limit. See Fig.~\ref{Fig:Star} for a numerical integration of the semiclassical equations with the MOR-RSET. The principal qualities of these stars, such as a wide interior region of negative Misner-Sharp mass and classical pressures that grow inwards, are shared with the semiclassical stars found with the RP-RSET. Their associated mass-radius diagram is also qualitatively identical to Fig.~\ref{Fig:MtoR}. 

\subsubsection{Summary of the section and further comments}

We have followed two unrelated approaches to modelling the RSET of a massless, minimally coupled scalar field, whose backreaction allows, in both cases, for the existence of a new type of ultracompact object supported by quantum vacuum polarisation. The robustness of this result is manifested through its generality, as both approaches not only agree in the existence of semiclassical black stars, but also on their most basic properties. What makes their matter overcome gravitational contraction is the presence of a negative mass interior, generated by the negative energy densities characteristic of the vacuum. These negative masses exert an additional gravitational repulsion that supports the object, while
relation~\eqref{Eq:Cont} translates their inwards-increasing classical pressures into large interior redshifts. Semiclassical stars exhibit a photon sphere, so they could produce gravitational wave echoes. To give a rough estimate, for the star obtained through the MOR-RSET model from Fig~\ref{Fig:Star}, the crossing time for a null ray emitted from the photon sphere to be reflected at $r=0$ and return to the photon sphere is
one order of magnitude larger (in units of the stellar mass $M$) than the time it takes to go from the photon sphere to the surface, i.e.,
\begin{eqnarray}\label{Eq:CrossingTime}
    \tau_{\text{echo}}= 2 \int_{0}^{r_{\text{ph}}}\left(h/f\right)^{1/2}dr' \simeq 306 M \sim 52 \tau_{\text{s}},
\end{eqnarray}
where $\tau_{\text{s}}$ is the light-crossing time between the photon sphere and the star surface.
Attending only to the geometrical features, this effect could result in gravitational-wave echoes clearly distinguishable from the characteristic signals associated to BH mergers.  

These stars might be the end-state of gravitational collapse, if at some point after the initial collapsing phase the dynamics takes the configuration to situations allowing for the accumulation of significant vacuum polarisation. Such situations are in fact more plausible than one might think. Semiclassical effects may turn out to be capable of significantly altering the dynamics of trapping horizons on much shorter timescales than what was previously believed, altering the standard gravitational collapse paradigm by rendering generic trapped regions dynamically unstable. We explore this possibility, along with scenarios of collapse which tends to relax to staticity, in the following section.

\section{Dynamical situations beyond the Hawking evaporation paradigm}\label{Sec:Dynamical}

There is by now a long-established picture of BH formation and evolution in semiclassical gravity. It is widely regarded that the classical collapse to a BH has negligible short-term semiclassical corrections, with the final configuration only deviating from its classical behaviour in the subsequent, very slow process of Hawking evaporation~\cite{Hawking1975}. Within this picture, the static ultracompact solutions presented above would be of little physical interest, as they would not match with any part of such dynamical scenarios. Even if such objects could be formed when starting from some very particular initial conditions, if they were to be compact enough to mimic BHs, they would likely be dynamically unstable: perturbations around them would tend to form trapped surfaces which enclose them, independently of how the perturbation and object actually interact~\cite{Mathur2022}.

However, this can all change if the effective semiclassical evolution of generic trapped regions turns out to be different from what is commonly believed. Indeed, the only established fact is that once a trapped region is formed, the configuration quantum fields relax to leads to no significant short-term semiclassical corrections to the evolution of the geometry in the vicinity of the \emph{outer horizon}. From there, making use of classical intuitions~\cite{Penrose1964,HawkingEllis1973}, it is assumed that anything that happens inside the trapped region cannot affect the region external to the BH. The flaw in this logic is apparent: Hawking evaporation itself already breaks the causal principles which forbid contact between the interior and exterior of the BH. The RSET typically violates energy positivity conditions in the vicinity of horizons, rendering classical intuitions useless. Thus, in semiclassical gravity (and beyond), the external universe is not safe from what may transpire inside a BH, and a thorough analysis of the depths of these objects is called for.

The simplest BH model is the Schwarzschild solution, which is comprised of a spacelike singularity fully enclosed by a trapped region. Being the simplest BH, this solution has typically been used as a background in semiclassical toy models, due mainly to the notorious complexity of quantum field theory in less symmetric spacetimes (such as actual dynamical collapse scenarios). However, perhaps unsurprisingly, this type of toy model is not sufficient to get a full picture of semiclassical BH formation and evolution. On the one hand, the inner structure of generic classical BHs is quite different from the vacuum Schwarzschild geometry, generally having an inner horizon~\cite{Wald1984} (albeit a classically unstable one), as well as an approach toward a potentially oscillatory BKL singularity at their centre~\cite{BKL}. On the other hand, one finds that even the vicinity of the outer horizon can be subject to large semiclassical corrections, given the right dynamical circumstances (indeed, the very effects which sustain the ultracompact configurations seen previously).

In this final section of the present chapter, we will aim to convey two particular results which showcase the potential for deviations from the standard Hawking evaporation picture. The first result is the fact that a dynamically formed inner horizon suffers not only from a classical instability (mass inflation~\cite{PoissonIsrael89}), but also from a semiclassical one, which tends to have the opposite dynamical effect to its classical counterpart, reducing the size of the trapped region from the inside-out. The second result is the fact that, given an appropriate dynamical scenario, the ``in" vacuum state of gravitational collapse can have a behaviour in therms of its energy content similar to the divergent Boulware state, which sustains horizonless ultracompact objects.

\subsection{Inner horizon classical and semiclassical dynamics}

Semiclassical analyses of BH evolution typically focus on the vicinity of the outer apparent horizon of the trapped region. This being the standard definition of the ``size" of a BH, it would be a good enough criterion for the state of these objects if their interior were indeed causally separated from the external universe, as is the case classically. However, due to the fact that the RSET can violate energy positivity conditions~\cite{Visser1996}, semiclassical analyses must include a description of the interior of these objects, particularly focusing on the dynamics of the trapped region as a whole. Since making sense of the high-curvature region at the centre of these objects is typically considered to go beyond the scope of the semiclassical approach, one should focus first on the region where this approach is potentially still applicable and, as we will see, necessary to understand the full dynamics of BHs: the \textit{inner} apparent horizon.

The geometries of generic BHs, such as those with electric charge and/or angular momentum, posses an inner horizon. Classically, this horizon has been shown to be prone to dynamical instability. The so-called \emph{mass inflation} instability~\cite{PoissonIsrael89,Ori1991} is triggered when, after gravitational collapse, the newly formed BH with an inner horizon receives an ingoing flux of energy from the natural decay tails of fields in its vicinity. Semiclassically, a similar instability is triggered, sourced by flux terms present in the RSET of quantum fields~\cite{BalbinotPoisson1993,Hollands2020a,Zilbermanetal2020}. While initially the flux terms which source the classical instability are larger, their decay in time implies that the semiclassical instability may eventually dominate the dynamics. Indeed, analyses on fixed backgrounds (in the absence of backreaction) do suggest such a late-time semiclassical dominance. This fact has previously been used to argue in favour of the cosmic censorship conjecture~\cite{Hollands2020a,Zilbermanetal2020} due to the elimination of potential extendability of the geometry past its Cauchy horizon. However, recent works suggest that this semiclassical dominance could have a much more drastic effect on the dynamics of the BH spacetime, potentially eliminating the trapped region through an outward displacement of the inner horizon on short timescales~\cite{Barceloetal2020,Barceloetal2022}.

In the following we will review the main features of both the classical and semiclassical inner horizon instabilities, with the aim of comparing them and building a clear picture of the possible outcomes of the full dynamical evolution of BHs.

\subsubsection{Classical mass inflation instability}

A general characteristic of static and stationary BH geometries which posses an inner horizon is the presence of a Cauchy horizon, beyond which the evolution of spacetime can no longer be determined by initial conditions from a past (partial) Cauchy surface \cite{Wald1984,HawkingEllis1973}. The limit in which this latter horizon is formed is a useful tool to understand the origin of the instability of the inner horizon.

\begin{figure}
    \centering
    \includegraphics[scale=.8]{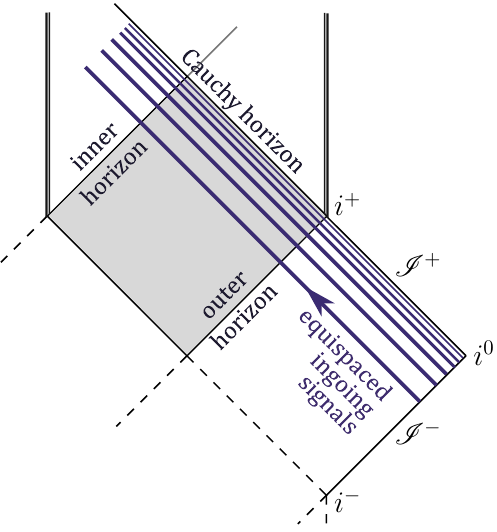}
    \caption{Causal diagram of the (future extension of) a BH with an inner and Cauchy horizons (the past must be connected with a matter collapse in a physical scenario). A series of signals is sent into the BH on regular time intervals, as measured by $v$. Due to the discrepancy between the infinite external time $v$ and the finite affine time it takes to cross the Cauchy horizon, the spacing between the signals, and hence their energy density, is infinitely blueshifted.}
    \label{f11}
\end{figure}

To get a general grasp of the features of such a geometry, let us look at a simple example. Consider the static spherically-symmetric metric, written in advanced Eddington-Finkelstein coordinates,
\begin{eqnarray}\label{metric-vr}
ds^2=-f(r)dv^2+2dvdr+r^2d\Omega^2.
\end{eqnarray}
This represents a BH with an inner and outer horizon if the redshift function $f(r)$ has two zeros, one with negative and one with positive slope. We can expand $f(r)$ around the inner horizon (the negative-slope root) $r_{\rm i}$, as
\begin{eqnarray}\label{fexp}
f(r)=-2\kappa_{\rm i}(r-r_{\rm i})+\order{(r-r_{\rm i})^2},
\end{eqnarray}
with $\kappa_{\rm i}>0$. The first important feature of the inner horizon is the accumulation of geodesics which approach it in time. Particularly, if we consider outgoing null geodesics, equating the line element \eqref{metric-vr} to zero and using \eqref{fexp} we find that the solutions for their radial motion in time is
\begin{eqnarray}
r-r_{\rm i}\propto e^{-\kappa_{\rm i} v}.
\end{eqnarray}
The second feature of note is the evolution of the affine parameter $\sigma$ along these geodesic trajectories. If we take one of the geodesic equations for radial motion $(v(\sigma),r(\sigma))$,
\begin{eqnarray}
\ddot{v}=-\frac{\partial_rf}{2}\dot{v}^2,
\end{eqnarray}
where the dot indicates a derivative with respect to the affine parameter $\sigma$, with the expansion~\eqref{fexp} we get that the relation between the advanced time $v$ and the affine parameter $\sigma$ of geodesics which remain in the vicinity of $r_{\rm i}$ is
\begin{eqnarray}\label{proptime}
\sigma\propto e^{-\kappa_{\rm i}v}.
\end{eqnarray}
For outgoing null geodesics which tend toward the inner horizon asymptotically, this relation becomes exact. The affine parameter of these geodesics therefore tends to a finite value when $v\to\infty$. In the absence of a singularity in this limit, a Cauchy horizon is formed, beyond which these geodesics, and the geometry itself, are extendable.

The relation~\eqref{proptime} between proper time near the Cauchy horizon and far away from the BH can be related with a blueshift amplification of perturbations~\cite{Simpson1973}, as can be seen schematically in Fig.~\ref{f11}. If one were to send lightlike signals into the BH at regular time intervals form the outside, then the spacing between those signal as perceived by an observer approaching the Cauchy horizon would tend to decrease to zero. Said observer would then get hit with infinitely many signals compacted in a nearly infinitesimally thin region. Even if the strength of each successive signal were to decrease, say, as an inverse polynomial in time (such that the total energy thrown into the BH is finite)~\cite{Price1971}, the exponential blueshift would still lead to a divergent energy density accumulation at the Cauchy horizon.

Aside from the headache this would likely cause the observer crossing this horizon, one might also expect that spacetime itself would react violently to such an energy accumulation. Indeed, when one considers backreaction on the geometry, one finds that the so-called \emph{mass inflation} instability is often triggered in such conditions.

While this argument gives a clear intuitive picture of the origin of mass inflation, before we describe the characteristics and consequences of this instability we must stress two important facts, which will be significant in later parts of this section. The first is that the blueshift described is a local effect produced in the vicinity of the inner horizon, and it is not dependant on the global causal structure of the spacetime being like the one depicted in Fig.~\ref{f11}. Mass inflation already has a substantial effect at finite values of the time coordinate $v$~\cite{Marolf2012}, implying that even spacetimes in which the trapped region has a finite lifetime are not exempt from it.

The second observation is the fact that not all inner horizons are prone to mass inflation. On the one hand, it is clear that the exponential blueshift argument presented above relies heavily on the surface gravity $\kappa_{\rm i}$ being non-zero. If the surface gravity were to be zero, i.e. if the expansion \eqref{fexp} started at second or higher order in $(r-r_{\rm i})$, then mass inflation would indeed be expected to not take place~\cite{Carballo-Rubio2022b}. On the other hand, even when the surface gravity is non-zero, there are some requirements on the dynamical response of the horizon to a given perturbation which must be satisfied in order for the instability to be triggered. The shell-based model in \cite{Barceloetal2022} provides a clear example of this: for a generic metric of the type \eqref{metric-vr}, the instability was only observed when the inner horizon position had an inverse-polynomial relation to the mass of the ingoing perturbations (or exponential with an exponent coefficient smaller than $\kappa_{\rm i}$). In other words, when the ingoing part of the perturbation alone leads to a displacement in time
\begin{eqnarray}
    \Delta r_i\gtrsim e^{-\kappa_{\rm i} v}.
\end{eqnarray}
This is satisfied, for instance, by a Reissner-Nordström BH with an ingoing perturbation governed by the Price law~\cite{Price1971}. Then, when one adds an (arbitrarily small) outgoing component to the perturbation inside the trapped region, the gravity-mediated exchange of mass between the ingoing and outgoing sectors, combined with the displacement of the inner horizon, triggers the non-linear amplification of curvature.

If a geometry does undergo the mass inflation process, then the Misner-Sharp mass (in spherical symmetry) at and below the initial position of the inner horizon tends to grow either exponentially in $v$~\cite{Ori1991}, or polynomially in $v$~\cite{Carballo-Rubio2021}. As a consequence, the inner apparent horizon tends to plummet toward the origin, and the formation of a strong curvature singularity is approached. For a Reissner-Nordström geometry and perturbations modelled by a series of ingoing null spherical shells of decreasing mass and a single outgoing null shell inside the trapped region~\cite{Barceloetal2022}, the geometry takes on the form
\begin{eqnarray}\label{geomassinfl}
ds^2=A(v)\left[-A(v)F(v,r)dv^2+2dvdr\right]+r^2d\Omega^2,
\end{eqnarray}
where $A(v)\propto e^{-\kappa_{\rm i}v}$, with $\kappa_{\rm i}$ being the initial (absolute) value of the inner horizon surface gravity, and $F(v,r)$ is the Reissner-Nordström redshift function, except with an exponentially growing mass term~\cite{Barceloetal2022}.

\begin{figure}
    \centering
    \includegraphics[scale=1.4]{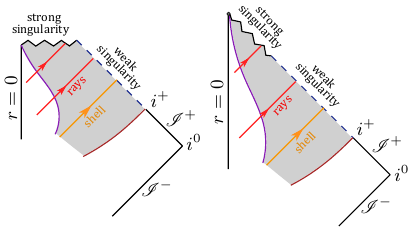}
    \caption{Future part of the causal diagram of a mass inflation geometry~\cite{Barceloetal2022}. The shaded part is the trapped region and the dashed line is the Cauchy horizon (and a weak singularity). The outgoing perturbation in the form of a shell is shown as well. Left: the inner apparent horizon reaches $r=0$ at finite $v$ and forms a Schwarzschild-type spacelike singularity. Right: the inner horizon only tends to $r=0$ asymptotically in $v$, resulting in a strong null singularity at $v\to\infty$ and $r=0$, at a finite affine distance for geodesics which fall into it.}
    \label{f12}
\end{figure}

The inner apparent horizon of the geometry~\eqref{geomassinfl} approaches the origin as $e^{-\kappa_{\rm i}v}$, but not all outgoing null geodesics in the trapped region approach this horizon. Due to the presence of the function $A(v)$, the movement of geodesics inside the mass-inflated region is quickly frozen, and many reach the limit $v\to\infty$ while still far away from the inner horizon and the origin. There are thus two separate regions these geodesics can end up in: some do indeed approach the inner horizon and the origin, and end up falling into a strong curvature singularity, while others are frozen at a finite radius and reach a Cauchy horizon. This latter horizon is now of variable radius, extending from the initial radius of the inner horizon all the way down to the origin. Additionally, due to the exponentially growing mass in this region, this horizon is also a singular surface. However, it still constitutes a Cauchy horizon because it turns out to be only \emph{weakly} singular, in the sense that tidal deformations across it are finite, and the metric is continuously extendable beyond it~\cite{Ori1991,Dafermos2017}. Causal diagrams of two possible such outcomes are depicted in Fig.~\ref{f12}.

\subsubsection{Semiclassical backreaction on static inner horizons}

Even before the classical mass inflation process was fully understood, the possibility of semiclassical effects destabilising such configurations was also contemplated. Unsurprisingly, the blueshift effect discussed above also affects the modes which define a field quantisation. The result is that regular quantum states defined from appropriate initial conditions in the past end up developing a singular behaviour at the Cauchy horizon. This is most clearly encoded in the RSET, which tends to a divergence there.

Following the spirit of classical arguments, the first semiclassical analysis of the problem~\cite{BirrellDavies1978} was formulated with the purpose of showing the impossibility of extending a universe through a Cauchy horizon, due to divergent values of the RSET backreacting on said horizon to create a singularity. Subsequent analyses~\cite{BalbinotPoisson1993,Zilbermanetal2020,Zilbermanetal2022,Hollands2020a,Hollands2020b} have continued in the same direction: calculating the RSET on a fixed static or stationary background (usually Reissner-Nordström or Kerr, with either flat or de Sitter asymptotics) in a physically reasonable vacuum state, and showing the divergent values of the RSET at the Cauchy horizon. The computational feat of performing such calculations in 3+1 dimensional backgrounds is tremendously impressive, and it is indeed clear by now that if a trapped region were to exist long enough for the formation of a Cauchy horizon to be approached, semiclassical backreaction would step in prevent it from happening.

However, it is worth remembering that semiclassically, trapped regions are not expected to last that long. So long as the outer horizon has a non-zero surface gravity, Hawking evaporation~\cite{Hawking1975} is expected to reduce the size of the BH from the outside, making its lifetime long but finite. Additionally, much like how classical mass inflation changes the size of the trapped region at finite times, making the inner horizon move inwards, one may expect that whatever the energy content in the RSET is which tends to a divergence might also cause some significant backreaction long before a Cauchy horizon is formed. This does indeed appear to be the case, and it may be quite significant for our understanding of BH evolution.

The divergence in the RSET is contained in the $\expval{T_{vv}}$ flux term, where $v$ is again the Eddington-Finkelstein ingoing null coordinate. The similarity with the classical case is apparent, since it is once again an ingoing flux which gets blueshifted infinitely. However, there are two crucial differences, as can be seen in e.g.~\cite{Hollands2020a,Zilbermanetal2020}. First, on a fixed background, the term $\expval{T_{vv}}$ becomes a constant when nearing the Cauchy horizon, unlike the classical perturbing flux which, tends to zero. Note that the $v$ coordinate is singular at this horizon \eqref{proptime}, making this a clear physical divergence. Second, the term seems to be \emph{negative}. The RSET often violates energy conditions~\cite{Visser1996}, so this is hardly a surprise. The important point is what this could imply for backreaction at finite times. Much like how a negative ingoing flux drives Hawking evaporation of the trapped region from the outside~\cite{DFU}, it is quite likely that the flux at the inner horizon would lead to classically forbidden dynamics, and the instability of this horizon could exponentiate the timescale involved.

\begin{figure}
    \centering
    \includegraphics[scale=1]{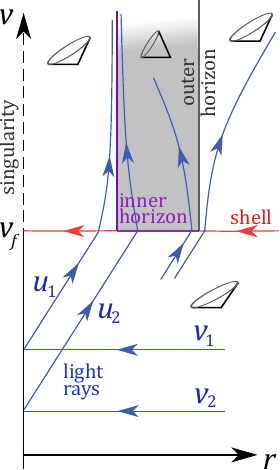}
    \caption{Spacetime diagram of the formation of a BH with an outer and inner horizon from the collapse of a null shell. The coordinates are the Eddington-Finkelstein $\{v,r\}$, and the shaded part is the trapped region.}
    \label{fig3}
\end{figure}

This is precisely what is found in \cite{Barceloetal2020}. The model considered there is quite simple: a BH with a geometry of the type \eqref{metric-vr} with an outer and inner horizon, formed from the collapse of a null spherical shell, as depicted in Fig.~\ref{fig3}. The shell is used to simplify the initial conditions for the quantum modes in the BH region for the ``in" vacuum state, and backreaction is analysed perturbatively from the point at which the inner horizon is formed. The RSET is calculated in the Polyakov approximation, and it has the expected negative ingoing flux at the inner horizon
\begin{eqnarray}
\expval{T_{vv}}|_{r=r_{\rm i}}=-\frac{l_{\rm p}^2\kappa_{\rm i}^2}{192\pi^2r^2_{\rm i}},
\end{eqnarray}
where again $\kappa_{\rm i}$ is (the absolute value of) the surface gravity of this horizon, $r_{\rm i}$ is its radial position, and $l_p$ is the Planck length.

A perturbation of the metric around the inner horizon (with a series expansion in $r$) sourced by the RSET leads to a modification of the geometry which tends to displace this horizon as
\begin{eqnarray}\label{ri}
r_{\rm i}\simeq r_{\rm i,0}+\frac{l_{\rm p}^2\kappa_{\rm i}}{48\pi\lambda}(e^{\lambda v/r_{\rm i,0}}-1),
\end{eqnarray}
with $\lambda=1+2\kappa_{\rm i}r_{\rm i,0}$, and assuming $v=0$ is the moment the semiclassical evolution begins. The position of this horizon changes in an outward direction, and the dynamical tendency has an initially exponential behaviour in time. While the approximation used to obtain this estimate breaks down after a small time interval, is suffices to show that backreaction is indeed likely to be amplified by the unstable nature of this horizon. If we extrapolate from this initial tendency and assume that the exponential behaviour persists throughout the evolution of the horizon, then one sees that the lifetime of the trapped region would be much shorter than what can be expected from Hawking evaporation from the outside. Particularly, assuming that the inner horizon surface gravity is initially larger than that of the outer horizon (which is typically the case), then the time it would take the perturbation to overcome its Planck-scale suppression and for the trapped region to be eliminated by the exponential outwards motion of the inner horizon \eqref{ri} would be
\begin{eqnarray}\label{vevap}
v_{\rm evap}\simeq\frac{1}{\kappa_{\rm i}\lambda}\log\frac{M}{l_{\rm p}}\lesssim\frac{M}{M_\odot}\times 10^{-5}\,\text{s},
\end{eqnarray}
where $M_\odot$ is the solar mass, and the logarithmic dependence has been omitted in the bound on the right-hand side due to the fact that for no astrophysically reasonable object would it increase the order of magnitude further.

This result, while being only an extrapolation of the initial tendency in a perturbative analysis, is already highly suggestive of the presence of an additional characteristic timescale in the semiclassical BH evolution problem which is much smaller than the notoriously large Hawking evaporation time~\cite{Hawking1974}. It also suggests that if BHs do disappear in this way, they may do so in a rather explosive manner. Such a phenomenon could potentially be astrophysically observable, and should it occur, its characteristics should be studied in detail once full solutions of semiclassical evolution are available. Full solutions would of course also incorporate at least a transient period of classical mass inflation, which we will discuss in the following.

\subsubsection{Full semiclassical evolution}

As we have seen, there are two dynamical effects which control the evolution of a BH inner horizon. On the one hand, there is classical mass inflation~\cite{PoissonIsrael89}, sourced by the inverse-polynomial decay of fields in a BH environment~\cite{Price1971}. On the other hand, there is the possibility of a semiclassical outward inflation of the inner horizon~\cite{Barceloetal2020}, sourced by the negative ingoing flux present in the RSET~\cite{BalbinotPoisson1993}. While the semiclassical effect is initially suppressed by a Planck constant, unlike its classical counterpart the flux which sources it does not decay in time, making it likely that its effect would be quicker to accumulate. In fact, the flux is proportional to the surface gravity of the inner horizon, suggesting that its value may even increase as mass inflation goes on. On these grounds, it would appear that the RSET ends up becoming the dominant source of dynamics, as was already argued in the analyses of the vicinity of the Cauchy horizon~\cite{BalbinotPoisson1993,Hollands2020a}. 

\begin{figure}
    \centering
    \includegraphics[scale=1.2]{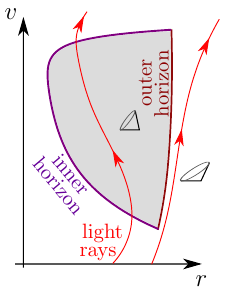}
    \caption{Spacetime diagram of the extrapolated dynamics of the semiclassical BH model obtained in~\cite{Barceloetal2022}. The inner horizon begins to move inwards due to mass inflation, but when the semiclassically sourced terms overcome the Planck scale suppression, it bounces back outwards and eventually meets the outer one.}
    \label{f3}
\end{figure}

It is therefore crucial to understand what finite-time backreaction form the RSET looks like on a background undergoing mass inflation, and seeing whether and when an outwards inflation of the inner horizon takes place. A first attempt at discerning this was made in \cite{Barceloetal2022}. A family of particularly simple geometries of the type \eqref{geomassinfl} which reproduce the causal properties of mass inflation was constructed (on purely geometric grounds), and backreaction from the RSET was solved for a small but finite time interval. What was found is that the motion of the inner horizon takes the form
\begin{eqnarray}
r_{\rm i}(v)=\alpha e^{-\kappa_{\rm i}v}+l_{\rm p}^2\beta e^{\kappa_{\rm i}v}+\cdots
\end{eqnarray}
with $\alpha$ and $\beta$ positive coefficients. The first term on the right-hand side is the classical trajectory of an inner horizon in the initial stages of mass inflation, and the second term is the correction sourced by the RSET. The approximation does break down when the correction becomes of the order of the background, but the tendency for an outward bounce is quite clear. Extrapolating from this initial tendency, an enticing possibility for the full semiclassical evolution of a BH emerges. As depicted in Fig.~\ref{f3}, if the exponential tendency form the correction term is continued, then the trapped region may disappear in a rapid outward burst, on a timescale similar to the estimate~\eqref{vevap}.

In light of these results, the possibilities for the full semiclassical evolution of the trapped region can be divided into two categories. First, while the observed initial tendencies for an outward horizon inflation are highly suggestive, we must reiterate that as of yet there are no full semiclassical solutions describing the complete evolution of a trapped region, even in the Polyakov approximation. It is therefore still possible that classical mass inflation prevails in the end, and curvature continues to increase until the depths of the trapped region can only be consistently described in a full quantum gravity theory. In such a scenario, if the quantum gravity effects do not lead to a quick bounce themselves, then the dominant dynamics would simply be the Hawking evaporation of the outer horizon.

The second category is the one more in tune with the observed tendencies: the RSET becomes the dominant source of dynamics at the inner horizon, and this horizon begins to move outwards. In this scenario, the two apparent horizons end up meeting (near the initial position of the outer one), and either a semiclassically stable extremal BH is left, or the trapped region disappears completely. From an observational standpoint, this last case would be the most intriguing. Not only would the disappearance of the trapped region likely be accompanied by a large burst of energy, which could potentially be observed, but a general instability of trapped regions would imply that the astrophysically observed long-lived dark ultracompact objects must in fact be horizonless BH mimickers, much like the ones described earlier in this chapter.

\subsection{Outer horizon and the vacuum state of gravitational collapse}

Even with the above analysis, there are still missing pieces in the picture which connects gravitational collapse with ultracompact objects as a final configuration. First of all, it is of course still premature to claim with certainty that long-lived trapped regions are not a stable and physically realisable configuration. Then, even assuming they are not (be it due to semiclassical or full quantum gravity effects), it is not fully clear how the transition between a collapse/bounce geometry and a static horizonless object may occur. Particularly, while in the previous section it is shown that vacuum energy can sustain ultracompact configurations with a surface arbitrarily close to the would-be horizon when the configuration is already static, it is not obvious whether and when such large vacuum energies can be achieved when starting from dynamical collapse.

What is missing is an understanding of what, if any, dynamical scenarios in which classical matter collapses can produce large vacuum energies which, in turn, can potentially stop the collapse. In other words, unlike the above analysis of inner horizon dynamics in which a small value of the RSET leads to a large dynamical contribution due to the instability of the geometry itself, here we would be looking for cases in which the RSET directly acquires large enough values to overcome its Planckian suppression and affect short-term dynamics. This was the subject of the works~\cite{Barceloetal2019,Barceloetal2020b}, and here we will briefly outline the results.

\subsubsection{Vacuum state of gravitational collapse}

Gravitational collapse geometries are typically constructed as an isolated system in an asymptotically flat spacetime. This not only simplifies the classical treatment of the system, but also the semiclassical one. Particularly, having an asymptotically flat region in the past (without horizons) gives an unambiguous choice for a quantum vacuum state: the Minkowski vacuum, constructed from the plane-wave mode solutions for the free field. Evolving this state from the past flat region into the future collapse geometry, one obtains the so-called ``in" state, which is the standard choice in collapse scenarios.

When a trapped region forms, the ``in" state quickly relaxes to the so-called Unruh state around and above the outer horizon~\cite{FabbriNavarro-Salas2005}. This state contains the fluxes governing the Hawking effect, but its energy content is small enough to not produce any significant short-term backreaction in the vicinity of the outer horizon. Additionally, past analyses have established that in a standard collapse, in which matter generally moves inwards quite quickly, there are no significant vacuum energy contributions close to the formation of a trapped region even before the relaxation to the Unruh state~\cite{DFU,ParentaniPiran1994,Barceloetal2008,Arderucio-Costa2017}.

However, the ``in" state can also have a different behaviour: if instead of forming a BH the matter content stabilises to a static configuration, then the ``in" state relaxes to the so-called Boulware state~\cite{Boulware1974}. This state is known for its divergence when defined on a BH spacetime, but it is perfectly well defined, and is in fact the physically reasonable choice, in stellar configurations. If the stellar object has a compactness close to that of a BH, then the vacuum energy does indeed approach a divergence, which turns out to work in favour of stabilising the configuration, as discussed in the previous chapter.

With this in mind, it becomes apparent that the the dynamical collapse scenarios most likely to produce large vacuum energies (in the ``in" state) are ones in which the matter content satisfies appropriate conditions on both its position and velocity: it must be both close to forming a horizon, and close to staticity. In such regimes, one may expect that the ``in" vacuum could have a behaviour somewhere in between the tendencies toward the Unruh and Boulware states, and this is indeed what is found.

\subsubsection{Dynamical approach to horizon formation: a thin shell model}

In \cite{Barceloetal2019} (followed up by~\cite{Barceloetal2020b}) we analysed a series of simplified collapse scenarios which approach the above conditions (staticity and horizon formation) in different ways. The first model we looked at was that of a thin spherical shell oscillating just above the Schwarzschild radius corresponding to its mass, paying particular attention to the part of its dynamics where it decelerates from collapse, halts and bounces back out. The second model was one in which a shell crosses its Schwarzschild radius, but does so at an arbitrarily low speed. The third was a case in which the formation of a single (marginally) trapped surface is only approached asymptotically in time. Let us briefly outline the results.

For the first two scenarios, the thin shell model simply consists of a Schwarzschild exterior and a Minkowski interior, matched together with appropriate junction conditions (which ensure the continuity of the metric) on a surface of variable radius $R(v)$ which we are free to fix. In double null coordinates the line element of this geometry is
\begin{eqnarray}
ds^2=\begin{cases}
-f\,du_+dv_++r^2d\Omega^2\quad\text{for}\quad r>R(v),\\
-du_-dv_-+r^2d\Omega^2\quad\text{for}\quad r<R(v),
\end{cases}
\end{eqnarray}
where the redshift function is the standard $f=1-2M/r$, with $M$ the shell mass. The evolution of the ``in" state is determined by the relations between the pairs of null coordinates at the shell surface, which we define as the functions
\begin{eqnarray}
g=\frac{du_+}{du_-}=\left.\frac{1}{\sqrt{|f|}}\sqrt{\frac{\alpha_-}{\alpha_+}}\right|_{\rm shell},\qquad h=\frac{dv_+}{dv_-}=\left.\frac{1}{\sqrt{|f|}}\sqrt{\frac{\alpha_+}{\alpha_-}}\right|_{\rm shell},
\end{eqnarray}
where the parameters
\begin{eqnarray}
\alpha_-\equiv\left.\frac{dv_-}{du_-}\right|_{\rm shell},\qquad \alpha_+\equiv\left.\frac{dv_+}{du_+}\right|_{\rm shell}
\end{eqnarray}
represent the velocity of the shell as seen from the inside and outside, in terms of the slope of its trajectory in the light cone ($\alpha_\pm=0$ and $\alpha_\pm=\infty$ correspond to ingoing and outgoing light speed, respectively, while $\alpha_\pm=1$ corresponds to staticity). The RSET is determined by the behaviour of these functions and their derivatives, as discussed in detail in~\cite{Barceloetal2019}.

\begin{figure}
    \centering
    \includegraphics[scale=.6]{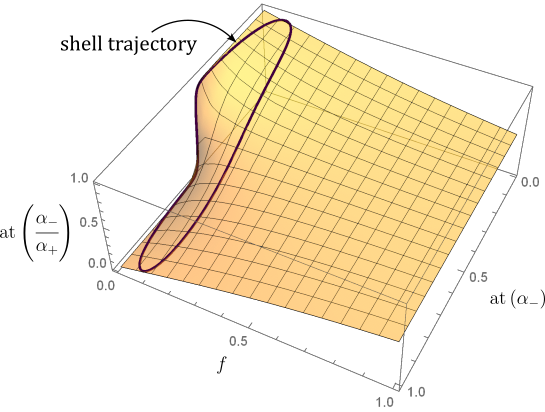}
    \caption{Velocity-dependent term of the relations $g$ and $h$ between the null coordinates on each side of the shell. The horizontal axes are the redshift function $f$ (as a compactification of the radial coordinate) and the velocity from the inside $\alpha_-$. Both the vertical and $\alpha_-$ axes are rescaled by $\text{at}(x)=\frac{2}{\pi}\text{arctan}(x)$. The closed curve shows an oscillatory shell trajectory.}
    \label{figDoppler}
\end{figure}

The main conclusion from this analysis comes form the fact that the functions $g$ and $h$ both have a dependence on two distinct terms: a redshift-dependent term, and a velocity ($\alpha_\pm$) dependent term. The velocity dependence in particular is the novel result, as it turns out that it can in fact potentiate the large vacuum energy production when matter slows down just before horizon formation. As shown in Fig.~\ref{figDoppler}, the slope of the quotient $\alpha_-/\alpha_+$, which is part of what determines the magnitude of the RSET, diverges at the point of zero velocity and zero redshift (the horizon position). In other words, if matter finds itself collapsing very slowly close to the formation of a horizon, then the vacuum energy produced is not only that given by a quasi-stationary approach toward the Boulware state, but is in fact larger due to the approach toward an extra divergence in the velocity terms.

The consequence of this for the oscillating shell is a production of large bursts of radiation whenever its trajectory passes through the region of large gradient of Fig.~\ref{figDoppler}. If, for instance, this model were to represent a perturbed ultracompact object, then perhaps these bursts could serve to quickly dissipate the perturbation, and they may potentially be observable form a great distance. However, a complete model of dynamical BH mimickers would be necessary to establish the scale involved in a physical realisation of such a scenario.

The second model we analysed was the case in which the shell actually crosses its Schwarzschild radius, but at a low speed (with $\alpha_\pm\sim 1$). What we found is that at and around the outer horizon, just after the shell crosses it, the RSET can acquire vary large values, the magnitude of which is qualitatively given by
\begin{eqnarray}
\expval{T_{\mu\nu}}\sim\frac{1}{(1-\alpha_-)^4}.
\end{eqnarray}
After this, however, the ``in" state predictably relaxes to the Unruh state and the large RSET values decay to the small energy content of this state. As this decay happens exponentially, the time it takes for the BH to ``thermalise" is given by
\begin{eqnarray}
v_{\rm therm}= 8M\log(\frac{1}{1-\alpha_-}).
\end{eqnarray}

The final dynamical model we looked at (and developed more thoroughly in~\cite{Barceloetal2020b}) is one in which the formation of a trapped surface is approached asymptotically in time. For this analysis we needed to go beyond the thin shell model. To summarise the results briefly, we found that in such cases the ``in" state can have a hybrid behaviour between the Boulware and Unruh states: it can produce a thermal emission of radiation at infinity, but at a temperature which is lower than the surface gravity of the asymptotically formed horizon. This mismatch in temperature then leads to the approach toward a Boulware-like divergence in the vacuum energy at the asymptotic horizon. The implication this may have for horizonless ultracompact objects is the fact that they can potentially have phases of thermal Hawking-like emission, while also retaining the large vacuum polarisation which sustains them.

\section{Summary and conclusions}

In this contribution we have discussed the potential effects that considering a physical vacuum, as opposed to the standard empty space of classical GR, could have on the dynamics of gravitational collapse and its end-point configurations. Considering the presence of quantum fields, as long as there is some spacetime curvature, even regions without classical matter should contain some average energy and pressures. This implies the presence of a new source term in the Einstein equations, which, being ultimately a function of the geometry itself, can alternatively be seen as a modification of the theory of gravity. One can picture the physical vacuum as a dielectric which can be polarised and, in extreme situations, even broken, giving rise to the creation of real particles.

Our analysis was restricted to spherically-symmetric situations, and we made use of qualitative approximations to vacuum effects based on simple analytic approximations to the Renormalised Stress-Energy Tensor (RSET) of a simple scalar field. Particularly, we used the (regularised) Polyakov approximation, and the Order-Reduced Anderson-Hiscock-Samuel approximation. We have recovered the standard Hawking result of BH evaporation of the outer horizon, and we have gone beyond it, identifying and analysing new situations which have previously not received much attention in the literature.

On the side of our analysis pertaining to static configurations, we have asked ourselves what effect a physical reactive vacuum can have on the fine balance of stellar equilibrium. After familiarising ourselves with the effects of the physical vacuum, we have searched for self-consistent regular static solutions of the stellar equilibrium equations for the simple case of constant-density stars. By self-consistent we mean that the vacuum polarisation effects generated by the geometry is the vacuum part of the source that generates this very geometry. In technical terms, in these analyses we have used the Boulware vacuum for the field as it is the only genuinely static vacuum compatible with asymptotic flatness. Our philosophy in these analyses is restricted to see what types of static configurations could exist, and not whether Nature could reach them or not and how. 

We have found two families of solutions: a) Stars with sub-Buchdahl compactness, which are just slight perturbations of the classical Schwarzschild stars; b) Stars with super-Buchdahl compactness, which are absent in classical GR. The existence of this second group of stars is one of the main results of these analyses. Ultracompact stars (i.e. stars with compactness near that of BHs) are able to exist because vacuum polarisation generates a central core of negative energy able to support their structure against gravitational collapse (see Figure~\ref{Fig:Star}). We have found similar ultracompact configurations through the use of our two rather different approximations; in both cases they exhibit the same qualitative features. 

There is an extended belief that the dark and compact objects we know exist in the cosmos are GR BHs. This believe is based, on the one hand, on the fact that the BH model accommodates very well the observations, but also on the theoretical idea that no stellar equilibrium can exist beyond certain compactness. Here we have shown that this second idea is not as strong as typically assumed. Our analyses show that the ``astrophysical BHs'' that we observe might equally well be ultracompact stars than GR BHs.

On the side of our analysis pertaining to dynamical scenarios of gravitational collapse, we have analysed the effect of vacuum energy in scenarios which go beyond the standard Hawking evaporation paradigm. This paradigm rest on two essential pillars, one of which is indeed justified by quantitative estimates, while the other turns out to be on rather shaky ground, bringing the whole picture into question. The first pillar is the fact that in standard situations of stellar collapse, one finds no reason why matter should qualitatively change its behaviour when approaching horizon formation; from a local point of view matter does not even know that globally a horizon is approached and eventually formed. This is indeed the case physically, except if matter happens to linger near the point of horizon formation long enough for light to cross from one end of it to the other; then, non-local effects which relate to the global configuration could play a role, such as the potential approach of the dynamical vacuum state toward a behaviour more akin to the static Boulware state. However, as a typical collapse is assumed to be initiated from not very compact configurations (as compared to the BH limit), matter is expected to cross the horizon threshold at very high speeds. Under this high-speed condition one can show that semiclassical effects around the point of horizon formation are negligible and so are not necessary for analysing the dynamics of the collapse~\cite{CandelasHoward1984,Barceloetal2008,Barceloetal2019}. The second pillar is the assumption that after the outer horizon is formed, whatever it is that occurs in the depths of the BH is irrelevant for its short-term evolution as seen from the outside. This is an extension of the logic originating from the classical case, where energy positivity conditions are satisfied, but with the addition of the slow process of Hawking evaporation~\cite{Hawking1974}; this process, however, only depends on the properties of the geometry at and above the outer horizon, and only brings about a negligible amount of contact between the interior and exterior of the trapped region due to the motion of this horizon, making it nearly compatible with the classical logic.

Our analysis brings strong evidence against the robustness of the second pillar, which in turn eventually makes the assumptions behind the first one crumble as well. First, we consider the more realistic situation in which the trapped region formed after gravitational collapse does not only have an outer horizon, but also an inner horizon. Much like how the inner horizon is unstable on a classical level (mass inflation), we find that semiclassical perturbations tend to destabilise it as well. We begin by analysing the effect of the RSET-sourced perturbation alone, finding that backreaction leads to an initially exponential tendency for this horizon to move outwards, shrinking the trapped region from the inside. If we extrapolate from this initial tendency, then instead of mass inflation we would have an inner horizon outwards inflation. This effect of course has to contend with its classical counterpart, and where the overall inner horizon instability leads to is still an open question. However, we find that due to the decreasing nature of the classical perturbation source leading to mass inflation, and the constant or increasing semiclassical source, it is likely that the semiclassical effect wins in the end. We observed this in one particular toy model which reproduces the causal properties of a classical mass inflation background, but a complete analysis of a full physical scenario is still pending. If the inner horizon does inflate outwards, then it would meet the outer one on a very short timescale (milliseconds for a Solar mass object), much shorter than the Hawking evaporation time. The analysis of what may come after such an evolution enters the realm of speculation, though given the limited number of possibilities, we can make some educated guesses. For instance, one could form an extremal BH, or end up eliminating the two horizons altogether. As extremal BHs are known to be unstable classically~\cite{Aretakis2012}, it is tempting to consider this latter possibility as the most likely outcome, though the possibility of semiclassical effects recuperating the stability of extremal objects is not out of the question.

If the trapped region does disappear completely, if all this occurs before the interior has reached Planckian densities (ensuring the validity of the semiclassical approach), then the previously collapsing matter would be freed form its causal restriction and likely begin expanding outwards following the inner horizon.\footnote{Let us also mention that a bounce of the collapsing matter has also been proposed previously on different grounds, with a similar overall effect on the dynamics, though also including a transient anti-trapped region~\cite{Barceloetal2011,Barceloetal2015,Barceloetal2016}} Part of the energy of the initial configuration would be dissipated to infinity after such a process, partly through particle production from the vacuum, but also likely due to processes within the effectively classical matter. The leftover matter would then be left in a condition to undergo collapse for a second time. If the trapped region instability on short timescales is indeed generic, then this process would repeat until at some point, potentially after a number of cycles of bouncing, dissipation and recollapse, the situation will change in the sense that the hypothesis than matter approaches horizon formation while moving rapidly will not longer be justifiable. Thus, the first pillar supporting the Hawking paradigm will also come into question, opening the door to a myriad of different possible dynamical continuations of the process. Here, our analysis of low-speed approaches toward horizon formation can provide some insight. Particularly, we have shown that large vacuum energy is generated by such dynamics, larger in fact than what might be expected from a mere transition in behaviour of the ``in" vacuum towards a Boulware-like behaviour. We have seen that the slow but present dynamics potentiate this growth of energy even further, making it more likely that semiclassical effects result in considerable deviations from the classical collapse dynamics. Additionally, we have also shown that cases in which the formation of a horizon is only approached, or even reversed periodically, can lead to the emission of Hawking-like radiation at infinity, leading to further dissipation until a stable endpoint is reached.

These results point towards the possibility that the final result of such dynamics would likely be an approach toward staticity, in which the ``in" vacuum would end up with a full Boulware behaviour. In such a scenario, classical and semiclassical matter and energy sources would end up in one of the configurations of hydrostatic equilibrium analysed above. The analyses of near-horizon dynamics also point to the potential behaviour of the vacuum when such objects undergo perturbations: they can emit both a nearly thermal spectrum of Hawking-like radiation, and bursts with much larger amplitudes, the latter being potentially observable if the object is sufficiently compact.

Overall, these results point towards the existence of an alternative theoretical paradigm for the ultimate fate of gravitational collapse. They also serve as a reminder of the fact that the standard paradigm is also to very large extent theoretical, relying on a series of not-so-obvious assumptions. It is therefore clearly necessary to continue working on the phenomenology of all these possibilities, preparing the grounds for their eventual verification or falsification through observation. From the perspective advocated in this contribution, GR BHs may just be idealised models of much more complicated stellar-like bodies.

\section*{ACKNOWLEDGEMENT}
The authors would like to thank Raúl Carballo-Rubio and Luis J. Garay for enlightening discussions and for all their work in our multiple collaborations, without which this contribution would not have been possible. 
Financial support was provided by the Spanish Government through the projects PID2020-118159GB-C43, PID2020-118159GB-C44 (with FEDER contribution), and by the Junta de Andaluc\'{\i}a through the project FQM219. 
J.A. and C.B. acknowledge financial support from the grant CEX2021-001131-S funded by MCIN/AEI/ 10.13039/501100011033.
V.B. acknowledges financial support provided under the European Union’s H2020 ERC Advanced Grant “Black holes: gravitational engines of discovery” grant agreement no. Gravitas–101052587.



\end{document}